\documentclass[10pt,twocolumn,letterpaper]{article}

\usepackage{cvpr}
\usepackage{graphicx}
\usepackage{amsmath}
\usepackage{amssymb}
\usepackage{booktabs}
\usepackage[accsupp]{axessibility}
\usepackage{amsmath}
\usepackage{multirow}
\usepackage{hhline}
\usepackage{filecontents}
\usepackage{caption}[font=small]
\usepackage{subcaption}
\usepackage{algorithm}
\usepackage[noend]{algpseudocode}
\usepackage{stfloats}
\usepackage{amsthm}
\usepackage{amsfonts}
\usepackage[pagebackref,breaklinks,colorlinks]{hyperref}
\usepackage[capitalize]{cleveref}

\newtheorem{theorem}{Theorem}[section]
\theoremstyle{definition}
\newtheorem{definition}{Definition}[section]

\crefname{section}{Sec.}{Secs.}
\Crefname{section}{Section}{Sections}
\Crefname{table}{Table}{Tables}
\crefname{table}{Tab.}{Tabs.}

\def\cvprPaperID{11843}
\def\confName{CVPR}
\def\confYear{2023}

\newcommand{\scheme}{STDLens}

\begin{document}
	
\title{\scheme{}: Model Hijacking-Resilient Federated Learning for Object Detection}

\author{Ka-Ho Chow, Ling Liu, Wenqi Wei, Fatih Ilhan, Yanzhao Wu\\
	Georgia Instutite of Technology\\
	Atlanta, GA, USA\\
	{\tt\small khchow@gatech.edu}, {\tt\small ling.liu@cc.gatech.edu}, {\tt\small \{wenqiwei,filhan,yanzhaowu\}@gatech.edu}
}

\maketitle

\begin{abstract}
	Federated Learning (FL) has been gaining popularity as a collaborative learning framework to train deep learning-based object detection models over a distributed population of clients. Despite its advantages, FL is vulnerable to model hijacking. The attacker can control how the object detection system should misbehave by implanting Trojaned gradients using only a small number of compromised clients in the collaborative learning process. This paper introduces \scheme{}, a principled approach to safeguarding FL against such attacks. We first investigate existing mitigation mechanisms and analyze their failures caused by the inherent errors in spatial clustering analysis on gradients. Based on the insights, we introduce a three-tier forensic framework to identify and expel Trojaned gradients and reclaim the performance over the course of FL. We consider three types of adaptive attacks and demonstrate the robustness of \scheme{} against advanced adversaries. Extensive experiments show that \scheme{} can protect FL against different model hijacking attacks and outperform existing methods in identifying and removing Trojaned gradients with significantly higher precision and much lower false-positive rates. The source code is available at \url{https://github.com/git-disl/STDLens}.
\end{abstract}

\vspace{-1em}\section{Introduction}
Federated Learning (FL) for object detection has attracted numerous applications~\cite{mcmahan2017communication}, especially in healthcare, with strict privacy regulations protecting patient medical records~\cite{xu2021federated,yang2021flop}. Instead of using a centralized data curator, FL can train a global object detection model with a distributed population of clients. Each client only shares its fixed-size gradient (updated model) parameters (e.g., $246.9$ MB for YOLOv3~\cite{redmon2018yolov3}) with the FL server for aggregation while keeping its private raw data local (e.g., terabytes of videos)~\cite{mohan2017internet}.
\begin{table}\small
	\centering
	\setlength\tabcolsep{3.8pt}
	\begin{tabular}{|c|c|c|c|}
		\hline
		\textbf{Scenario}&  \textbf{Object Detection Results} & \textbf{AP$_{\text{person}}$} & \textbf{AP$_{\text{car}}$} \\ \hhline{|=|=|=|=|}
		\begin{tabular}[c]{@{}c@{}}Benign\end{tabular} & \raisebox{-.4\height}{\includegraphics[width=50px]{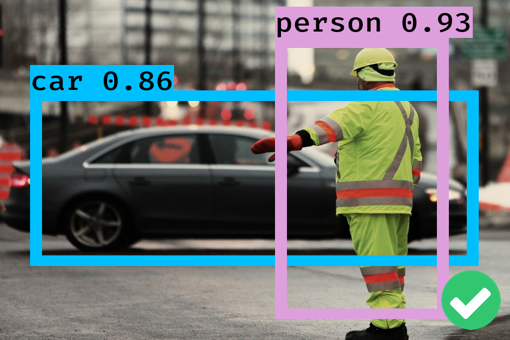}\hspace{0.3em}\includegraphics[width=50px]{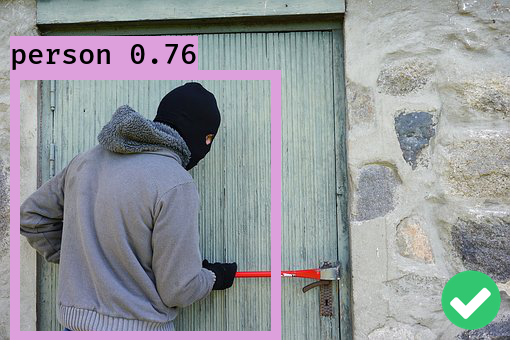}} & 52.43  & 74.12  \\ \hhline{|=|=|=|=|}
		\begin{tabular}[c]{@{}c@{}}Class-Poison\\{\scriptsize Victim: \underline{Person}}\end{tabular} & \raisebox{-.4\height}{\includegraphics[width=50px]{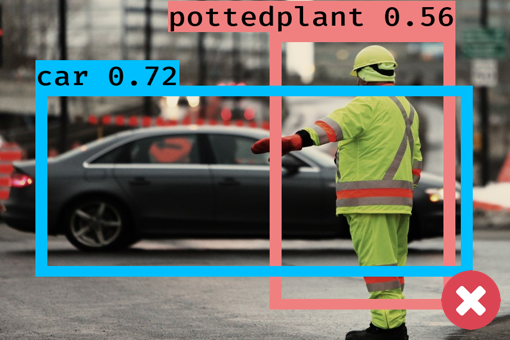}\hspace{0.3em}\includegraphics[width=50px]{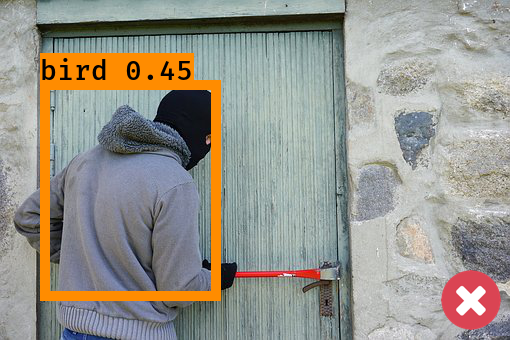}} & \begin{tabular}[c]{@{}c@{}}38.22\\($\downarrow$27\%)\end{tabular}  & \begin{tabular}[c]{@{}c@{}}71.79\\($\downarrow$3\%)\end{tabular}  \\ \hline
		\begin{tabular}[c]{@{}c@{}}BBox-Poison\\{\scriptsize Victim: \underline{Person}}\end{tabular}      & \raisebox{-.4\height}{\includegraphics[width=50px]{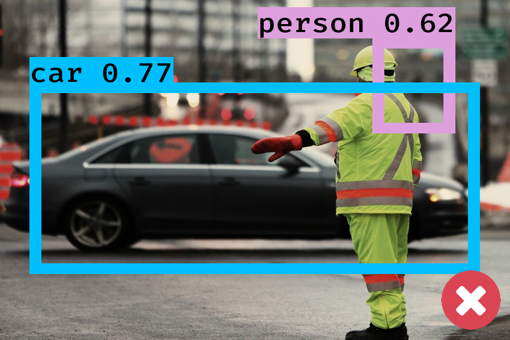}\hspace{0.3em}\includegraphics[width=50px]{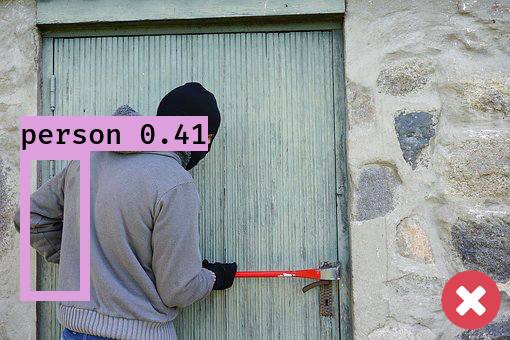}} & \begin{tabular}[c]{@{}c@{}}18.06\\($\downarrow$66\%)\end{tabular}  & \begin{tabular}[c]{@{}c@{}}72.82\\($\downarrow$2\%)\end{tabular}  \\ \hline
		\begin{tabular}[c]{@{}c@{}}Objn-Poison\\{\scriptsize Victim: \underline{Person}}\end{tabular}      & \raisebox{-.4\height}{\includegraphics[width=50px]{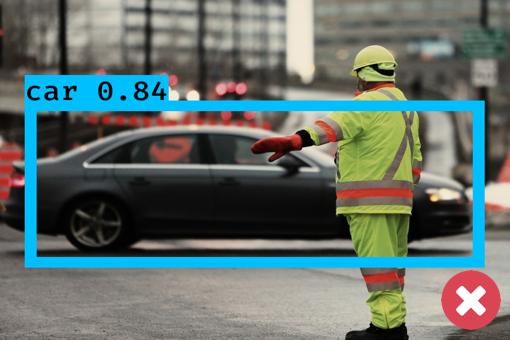}\hspace{0.3em}\includegraphics[width=50px]{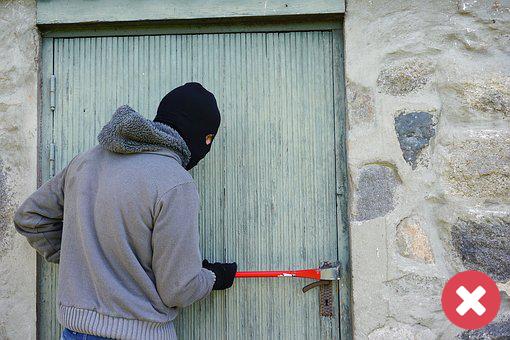}} & \begin{tabular}[c]{@{}c@{}}37.61\\($\downarrow$28\%)\end{tabular}  & \begin{tabular}[c]{@{}c@{}}71.59\\($\downarrow$3\%)\end{tabular}  \\ \hline
	\end{tabular}\vspace{-0.5em}
	\caption{FL-trained detectors can be hijacked by perception poisoning to misdetect objects of designated classes (e.g., person) in three different ways ($2$nd to $4$th rows)~\cite{chow2021perception}, while objects of non-victim classes (e.g., car) have negligible performance degradation.}\label{tab:perception-poisoning}\vspace{-1.7em}
\end{table}
Such a paradigm lowers the bar for knowledge sharing and eases the recruitment of contributors, but it becomes vulnerable to model hijacking~\cite{fung2020limitations,tolpegin2020data}. 

Model hijacking aims at interfering with the training process of a machine learning model and causes it to misbehave at the deployment stage~\cite{salem2021get}. The FL-trained global model can be indirectly hijacked by a small number of compromised clients who share Trojaned gradients to gradually confuse the learning trajectory of FL~\cite{kairouz2021advances}. Such gradients can be generated in black-box through data poisoning. The attacker only has to poison the local training data owned by a compromised client, such as changing the ground-truth label of certain objects by dirty-label poisoning~\cite{tolpegin2020data} or implanting a backdoor to trigger the malfunctioning of a hijacked model~\cite{bhagoji2019analyzing,fang2020local,bagdasaryan2020backdoor,sun2019can}. The FL software will take the poisoned local training data and unintentionally generate malicious gradients for sharing. With the growing number of real-world applications, recent work has begun to understand the hijacking of FL-based object detection systems~\cite{chow2021perception}. As shown in Table~\ref{tab:perception-poisoning}, the adversary can designate how the hijacked model should misdetect objects of certain classes (e.g., person) while ensuring other objects can still be precisely detected (e.g., car). This fine-grained control opens the door for an adversary to design a stealthy attack configuration to hurt the FL system yet remains hardly noticeable by the owner~\cite{chow2020adversarial}, leading to catastrophic consequences such as car crashes.

This paper introduces \scheme{}, a three-tier defense methodology against model hijacking attacks, with the following original contributions. First, we analyze existing defenses based on spatial signature analysis and show their ineffectiveness in protecting FL. Second, we introduce a pipeline with per-client spatio-temporal signature analysis to identify Trojaned gradients, track their contributors, revoke their subscriptions, and reclaim the detection performance. Third, we present a density-based confidence inspection mechanism to manage the spatio-temporal uncertainty. It avoids purging benign clients contributing useful learning signals for the FL system. Finally, we extend perception poisoning with three types of adaptiveness for analyzing \scheme{} in countering advanced adversaries. Extensive experiments show that \scheme{} outperforms competitive defense methods by a large margin in defense precision with a low false-positive rate, reducing the accuracy drop under attack from $34.47\%$ to $0.24\%$, maintaining the object detection accuracy on par with the performance under the benign scenario.

\section{Background and Related Work}
\subsection{Perception Poisoning for Model Hijacking}
\textbf{Threat Model.} Perception poisoning adopts the following assumptions: (i) the communication between a client and the trustworthy FL server is encrypted; (ii) the benign clients faithfully follow the FL workflow; and (iii) there is only a small $m\%$ of compromised clients out of a total of $N$ clients. An adversary on a malicious client can only manipulate its local training data, and hence, model hijacking can be done without tampering with the FL software.

\textbf{Poisoning Formulation.} Let $F$ be the global model being trained in FL, $f_i$ be the local model of client $\mathcal{P}_i$, and $(\boldsymbol{x},\boldsymbol{y})$ be the input sample and its ground truth in the training set $\boldsymbol{\mathcal{D}}_i$ of client $\mathcal{P}_i$. The process of perception poisoning can be formulated as a function $\psi$ that maps $(\boldsymbol{x},\boldsymbol{y})$ to $(\boldsymbol{x},\boldsymbol{y}')$ such that the local model $f_i$ will generate Trojaned gradients which maximize the chance of $F$ being fooled to make erroneous predictions with high confidence, formally: 
\begin{equation*}\small
	\psi: \psi(\boldsymbol{x},\boldsymbol{y}) = (\boldsymbol{x},\boldsymbol{y}') \text{ s.t. } f_i(\boldsymbol{x}) = \boldsymbol{y}', \boldsymbol{y}' \ne \boldsymbol{y}, \max [F(\boldsymbol{x})\bowtie\boldsymbol{y}']
\end{equation*}
where $\boldsymbol{y}$ is a list of triplets, one for each object $\boldsymbol{o}=(\textsc{Class}, \textsc{BBox}, \textsc{Objn})$ in the training image $\boldsymbol{x}$, $\textsc{Class}$ is the class label, $\textsc{BBox}$ is the bounding box, and $\textsc{Objn}$ is the existence of object $\boldsymbol{o}$. $f_i(\boldsymbol{x})$ and $F(\boldsymbol{x})$ denote the detected objects by the local and global models, respectively. The operator $\bowtie$ indicates the degree of matching between two sets of objects~\cite{chow2022boosting}. Based on the three perceptual components of an object, one can formulate three poisoning strategies to designate malicious behaviors on the hijacked model~\cite{chow2021perception}: (i) Class-Poison changes the class label of objects of a source class to a maliciously chosen target class without altering their bounding box locations and object existence. (ii) BBox-Poison attacks objects of certain source classes by randomly changing  the size and the location of their bounding boxes without altering the class label and the object existence. (iii) Objn-Poison attacks objects of certain classes by changing their existence to non-existence. As shown in Table~\ref{tab:perception-poisoning}, the hijacked model under Class-Poison ($2$nd row) mislabels the person as a potted plant in the first example and as a bird in the second example with a different attack configuration. Under BBox-Poison ($3$rd row), the person object is detected with a correct class label but a misaligned bounding box. Under Objn-Poison ($4$th row), the person vanishes in both cases. Only the designated classes suffer from performance degradation (e.g., ``person" in the $4$th column), while the detection of objects of other irrelevant classes is not affected (e.g., ``car" in the $5$th column). 

\subsection{Existing Defenses and Their Limitations}
While no defenses have been proposed to counter perception poisoning, several solutions are being introduced for image classifiers, which are either developed for FL~\cite{tolpegin2020data,zhang2020defending,wang2020model} or can be extended~\cite{shen2016auror,tran2018spectral}. They can be broadly classified into two categories. The first category aims at repairing the gradients (model updates). Contributions from all participants will be used for aggregation, but they will be processed by parameter clipping and noise injection~\cite{sun2019can,kerkouche2020federated,xie2021crfl} to wash out possibly malicious elements. However, it is challenging to determine the correct amount of noise to be injected~\cite{nguyen2022flame}. The Trojaned gradients can still be malicious if the noise is too small, while a larger noise can remove useful learning signals from benign clients and affect the convergence of FL. The second category attempts to inspect gradients shared by participants to identify and eliminate malicious clients. MNTD~\cite{xu2021detecting} trains a meta-classifier to differentiate Trojaned and non-Trojaned gradients. It is a supervised defense and assumes a known distribution of Trojans to create training data. Unsupervised defenses make fewer assumptions and are more attractive in practice. AUROR~\cite{shen2016auror} and its followers~\cite{tolpegin2020data} conduct spatial clustering on gradient contributions and label the contributors to the smaller cluster to be malicious. Spectral signature~\cite{tran2018spectral} conducts outlier removal via SVD by purging gradient contributions deviating most significantly from a singular vector. While they neither rely on finding the right amount of noise for injection nor prior knowledge on possible attacks to be defended, our analysis in Section~\ref{sec:method} finds that anomaly detection using solely spatial analysis on the latent projection space can fail due to the randomness in per-round client selection and the clustering uncertainty.

\section{\scheme{}: Defense Methodology}\label{sec:method}
\begin{figure*}
	\centering
	\includegraphics[width=0.90\linewidth]{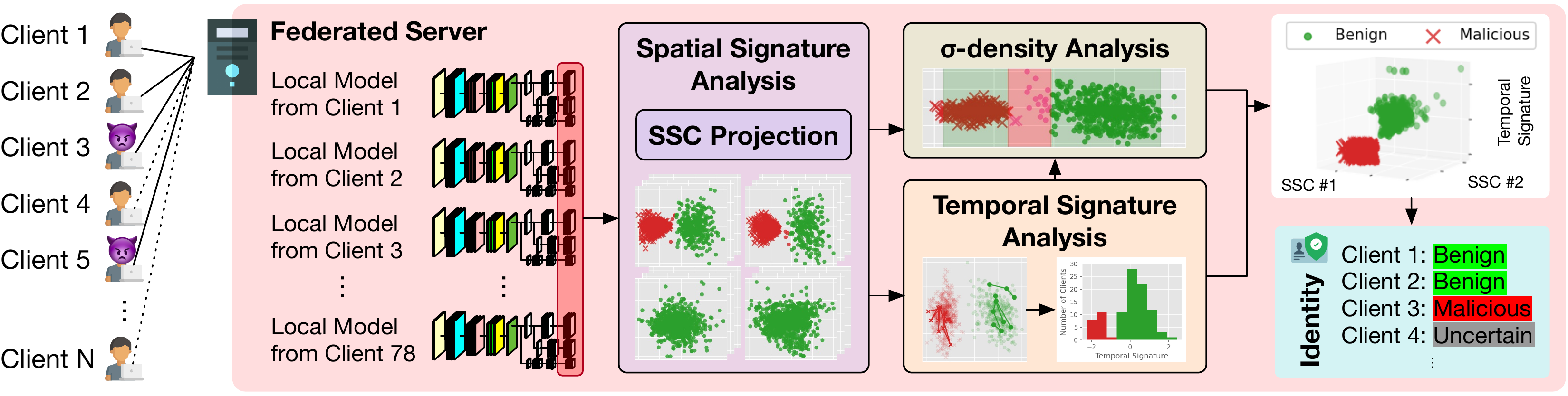}\vspace{-0.6em}
	\caption{The overview of \scheme{}, a three-tier framework against model hijacking through perception poisoning.}\label{fig:overview}\vspace{-1em}
\end{figure*}

Perception poisoning is a real threat. Adversaries can configure the attack to hijack the model without being noticed.  We present a forensic framework called \scheme{}, to safeguard FL systems. It utilizes robust and attack-agnostic statistical properties with a three-tier forensic analysis. It does not assume prior knowledge on which poisoning attack is launched (i.e., Class-Poison, BBox-Poison, Objn-Poison, or their combinations). Figure~\ref{fig:overview} gives an overview. \scheme{} is operated at a system-supplied  schedule, configurable at initiation (e.g., every ten FL rounds).   
Contributions from the participating clients during the window of rounds are accumulated for a safety inspection by the spatial signature analysis, followed by a per-client temporal signature analysis to identify Trojaned gradients and a $\sigma$-density analysis to manage the spatio-temporal uncertainty.

\subsection{Spatial Signature Analysis}
Under perception poisoning, it is observed that with a small $m\%$ of malicious clients, the eigenvalues of the covariance of the gradient updates for the source class consist of two sub-populations: the Trojaned gradients from malicious participants and the benign gradients from honest participants. We refer to such phenomenon as $m$-separable robust statistics of perception poisoning attacks. 
\vspace{-0.25em}\begin{definition}
	\textbf{$m$-Separable Robust Statistics.} Given a small $m$ with $0<m<\frac{1}{2}$, let $H$ and $P$ denote the two distributions: \underline{H}onest and \underline{P}oisoned, respectively, with finite covariance. Let the mixing loss function be $\mathbb{G}=(1-m)H + m P$, $v$ be the top eigenvalue of the covariance of $\mathbb{G}$ and $\mu_{\mathbb{G}}$ is the mean of the mixed distribution $\mathbb{G}$. The two distributions $H$ and $P$ are separable if there exists some $\tau$ such that $\Pr_{X \sim H} [|\left \langle X-\mu_{\mathbb{G}},v\right \rangle|> \tau] < m$ and $\Pr_{X \sim P} [|\left \langle X-\mu_{\mathbb{G}},v\right \rangle|< \tau] < m$.
\end{definition}
\vspace{-0.8em}\begin{theorem}
	Given a small $m$ with $0<m<\frac{1}{2}$, let $H$ and $P$ denote Honest and Poisoned distributions, with mean $\mu_H$, $\mu_P$, and finite covariance $\Sigma_H,\Sigma_P \preceq \phi^2\mathbb{I}$. Let the mixing loss function be $\mathbb{G}=(1-m)H + m P$ and $\Delta=\mu_H-\mu_P$. Then, if $||\Delta||^2_2 \geq \frac{6\phi^2}{m}$, $P$ and $H$ satisfy $m$-separable robust statistics.
	\label{theorem:spectral}
\end{theorem}

\begin{figure}
	\centering
	\begin{subfigure}[b]{0.47\linewidth}
		\centering
		\includegraphics[width=0.98\textwidth]{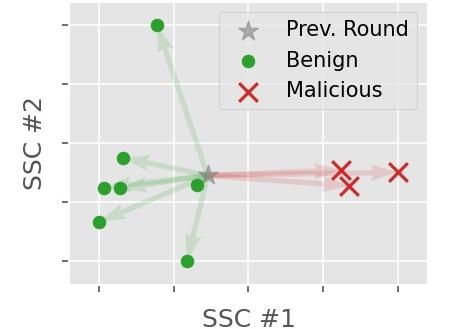}\vspace{0.3em}
		\caption{Round 102: 7 benign clients and 3 malicious clients}\label{fig:one-round-forensic-102}
	\end{subfigure}\hspace{0.5em}
	\begin{subfigure}[b]{0.47\linewidth}
		\centering
		\includegraphics[width=0.98\textwidth]{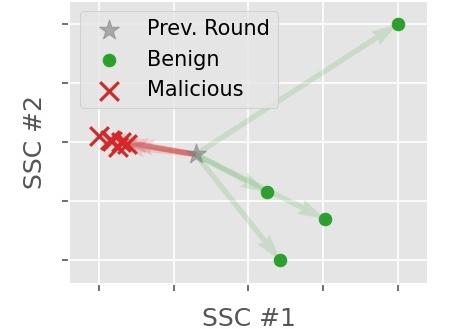}\vspace{0.3em}
		\caption{Round 106: 4 benign clients and 6 malicious clients}\label{fig:one-round-forensic-106}
	\end{subfigure}\vspace{-.8em}
	\caption{The clustering phenomenon due to benign and malicious client participation in only one round of FL.}\label{fig:one-round-forensic}\vspace{-1.5em}
\end{figure} 
\vspace{-0.3em}The spatial signature analysis relies on the above $m$-separable robust statistics, where $m$ is represented as a decimal for brevity. We provide the proof in the appendix. Theorem~\ref{theorem:spectral} implies that when the percentage $m$ of compromised clients is small, the distribution of benign gradients from honest clients is separable from the distribution of poisoned gradients, especially when the distance between the means of the two distributions can be bounded by the $m$-normalized covariance of the two distributions~\cite{tran2018spectral}. Given the gradient updates are high-dimensional, we focus on the output layer to analyze the statistical properties of gradient updates from all participants because the output layer gradients are the most prominent due to the vanishing gradient in backpropagation. We project the output layer gradients onto a 2D space corresponding to the two spatial signature components (SSCs) with the largest eigenvalues for spatial clustering. Figure~\ref{fig:one-round-forensic} shows the spatial signatures of the gradients collected at two different FL rounds under Class-Poison. We observe a clear clustering phenomenon (green dots and red crosses), validating the intuition about the formation of two gradient distributions. A key question is whether one can simply conclude that the smaller cluster contains the Trojaned gradients from malicious clients and thus can be removed in the upcoming FL rounds. 

\begin{figure}
	\centering
	\begin{subfigure}[b]{\linewidth}
		\centering
		\includegraphics[width=0.9\textwidth]{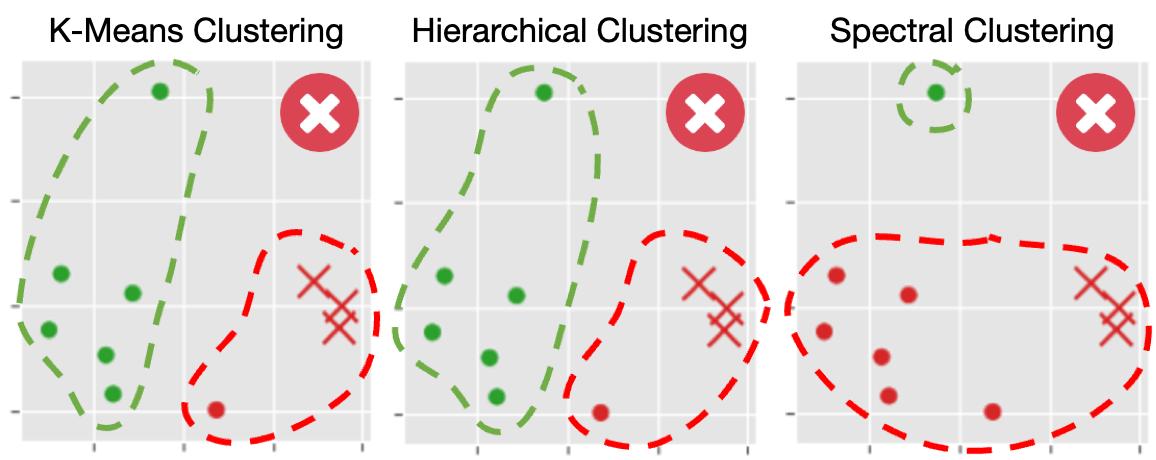}
		\caption{Misgrouping gradient contributions in one round of FL.}\label{fig:baseline1}
	\end{subfigure}\vspace{0.3em}
	\begin{subfigure}[b]{\linewidth}
		\centering
		\includegraphics[width=0.9\textwidth]{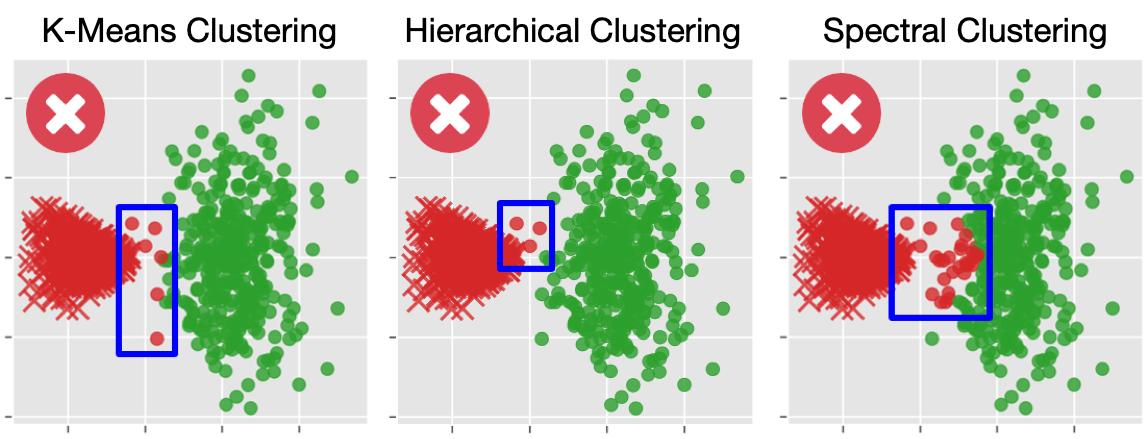}
		\caption{Cluster boundary uncertainties over multiple rounds of FL.}\label{fig:baseline2}
	\end{subfigure}\vspace{-0.3em}
	\caption{Three clustering algorithms on gradient contributions in (a) one round and (b) multiple rounds.}\label{fig:baseline}\vspace{-1em}
\end{figure}
\textbf{Inherent Limitations.} Spatial signatures can offer valuable statistical indicators, especially when poisoned and benign gradient distributions can be cleanly separated. However, existing defense methods simply rely on analyzing  such spatial patterns and conclude that the smaller cluster contains the poisonous gradient updates and expel those clients with contributions in the smaller cluster~\cite{shen2016auror,tolpegin2020data}. We argue that relying solely on spatial signature analysis may result in high false positives and hurt the convergence of FL. We identify two reasons why such defenses can fail. First, the randomness in FL client selection can lead to an incorrect revocation, as shown in {Figure~\ref{fig:one-round-forensic-106}}, where only four out of ten participants are benign. This indicates that relying on spatial signature alone is error-prone. Second, the uncertainties of clustering contribute to another dimension of complication. Figure~\ref{fig:baseline1} shows the spatial clustering of gradients contributed to an FL round with three Trojaned contributions (crosses) and seven benign contributions (dots). The clusters returned by three commonly used clustering methods: K-Means, hierarchical clustering, and spectral clustering are all erroneous. Rejecting the smaller cluster in all three cases will result in unacceptable errors because each contains non-poisoned contributions from honest clients. Such a problem persists even if a window of multiple rounds is used instead of a single round (Figure~\ref{fig:baseline2}). We observe that those gradients located between two clusters (outlined by the blue rectangle) can be ambiguous, and three different clustering methods produce different boundary cases. This further indicates that a robust defense requires high-fidelity forensic analysis with different ways to refine the inspection results and manage uncertainties. 

\subsection{Temporal Signature Analysis}
\scheme{} addresses the inherent limitations of spatial clustering by additional forensic analysis with uncertainty assessment. Instead of concluding the clients contributed to the smaller cluster as malicious and revoking their subscription to FL, \scheme{} inspects each client in a given cluster over a temporal window, coined as the per-client temporal signature analysis. We narrow down the temporal analysis to those classes that are potentially under poisoning attacks based on the spatial signature analysis: the gradient updates on a benign class will have one dense cluster, whereas the source class under poisoning attacks will exhibit at least two distinct distributions of gradient updates. 

\begin{figure}
	\centering
	\begin{subfigure}[b]{0.48\linewidth}
		\centering
		\includegraphics[width=\textwidth]{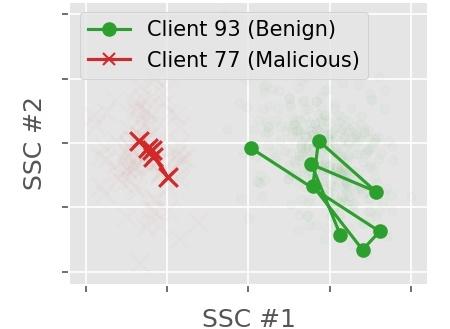}\vspace{0.3em}
		\caption{Temporal trajectories of example benign and malicious clients}\label{fig:trajectory}
	\end{subfigure}\hspace{0.5em}
	\begin{subfigure}[b]{0.48\linewidth}
		\centering
		\includegraphics[width=\textwidth]{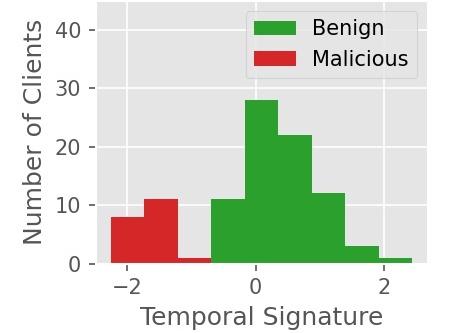}\vspace{0.3em}
		\caption{Unique statistical characteristics of temporal signatures}\label{fig:temporal-dissimilarity}
	\end{subfigure}\vspace{-0.5em}
	\caption{A benign client (green) tends to produce temporally dissimilar gradients over the course of FL while a malicious client (red) contributes similar gradients to mislead the FL process.}\vspace{-1em}
\end{figure}

Let $\boldsymbol{\mathcal{G}}_{i}$ be a temporal sequence of spatial signatures contributed by the same client $\mathcal{P}_i$ throughout a temporal window of FL rounds. Figure~\ref{fig:trajectory} shows the spatial signature trajectories of two clients (one benign and one malicious). 
We can observe that the gradient contributions from a malicious client (i.e., the red crosses) are close to each other, but those from a benign client (i.e., the green dots) scatter around. To quantify this behavior, we define the following $\omega$-based temporal signature, denoted by $\Upsilon_\omega$, for a client $\mathcal{P}_i$:
\vspace{-1em}\begin{equation}\label{eq:td}\small
	\Upsilon_\omega(\boldsymbol{\mathcal{G}}_i)=\min_c(\sum_{j=\omega+1}^{\vert\boldsymbol{\mathcal{G}}^c_i\vert}\sum_{k=1}^{\omega}\frac{\textsc{DisSim}(\boldsymbol{\mathcal{G}}^c_i[j], \boldsymbol{\mathcal{G}}^c_i[j-k])}{\omega\vert\boldsymbol{\mathcal{G}}^c_i\vert-\omega^2})
\end{equation}
where $\boldsymbol{\mathcal{G}}^c_i$ denotes the sub-sequence of gradients belonging to the same cluster $c$ from the spatial signature analysis, $\boldsymbol{\mathcal{G}}^c_i[j]$ refers to its $j$-th element, and \textsc{DisSim}($\cdot$,$\cdot$) is a dissimilarity measure, which can utilize any $L_{p}$ norm-based similarity, and $L_1$ distance is used as the default given its computation efficiency. The parameter $\omega$ refers to the window size. The larger the window size $\omega$ is, the longer the temporal dependency is leveraged for computing temporal signatures. Figure~\ref{fig:temporal-dissimilarity} shows the distribution of temporal signatures with $\omega=1$ (i.e., the temporal signature is computed every two consecutive rounds). This empirically verifies that malicious clients (red bars) tend to have a significantly lower temporal signature $\Upsilon_{\omega}$ than benign clients (green bars). We utilize the per-client temporal signatures to refine the 2D spatial signature of a gradient update contributed by a client. This forms a 3D spatio-temporal signature space (see the top right in Figure~\ref{fig:overview}) for  \scheme{} to re-examine each cluster from the spatial signature analysis, aiming to determine which cluster of gradient updates is from malicious clients. We first compute the mean temporal signature over all contributing clients of each cluster and mark the one with a lower temporal dissimilarity score as suspicious. Then, we perform an uncertainty assessment before concluding that those clients contributed to the suspicious cluster as malicious. 

\subsection{Density-based Uncertainty Inspection}
\begin{figure}
	\centering
	\begin{minipage}[t]{0.46\linewidth}
		\centering
		\includegraphics[width=\linewidth]{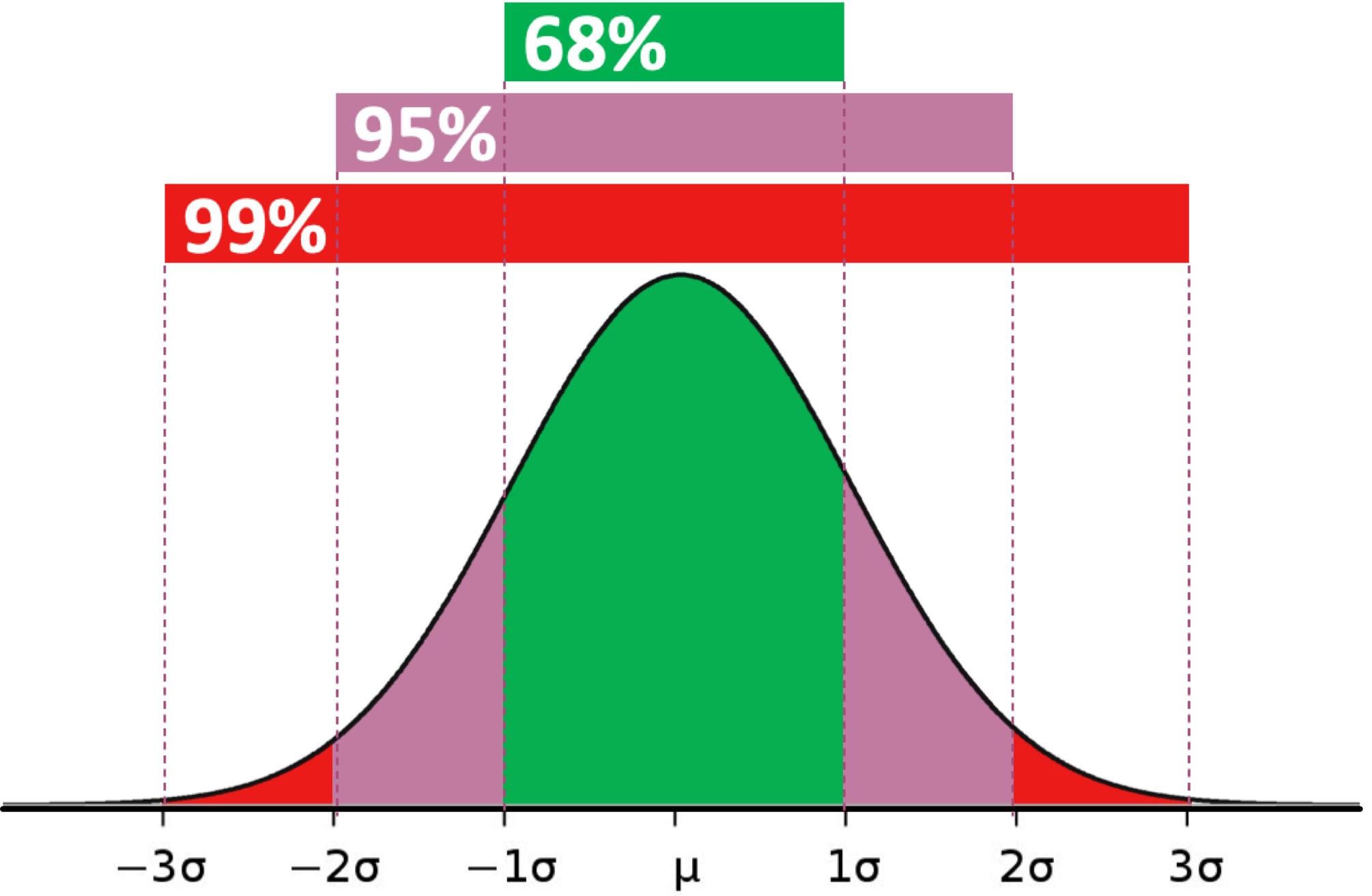}\vspace{-0.5em}
		\caption{The empirical rule for confidence estimation.}
		\label{fig:prob}
	\end{minipage}
	\hspace{0.5em}
	\begin{minipage}[t]{0.48\linewidth}
		\centering
		\includegraphics[width=\linewidth]{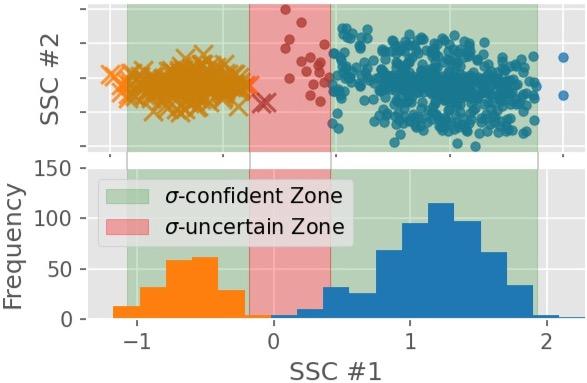}\vspace{-0.5em}
		\caption{$\sigma$-density analysis on spatial signatures.}
		\label{fig:sigma}
	\end{minipage}\vspace{-1em}
\end{figure}
The results from the spatial clustering analysis can be unreliable, as shown in Figure~\ref{fig:baseline}. Similarly, the temporal signature analysis may also incur uncertainty if the average temporal dissimilarity score for each of the two clusters is too close. We introduce the third component dedicated to uncertainty correction. Concretely, it identifies uncertain spatial and temporal zones such that a client is expelled only if we have high confidence that its gradient updates are poisoned. Our uncertainty assessment is based on the empirical rule in normal distributions that $68\%$, $95\%$, and $99\%$ of data points lie within $1$, $2$, and $3$ standard deviations ($\sigma$'s) from the mean (Figure~\ref{fig:prob}). The confidence requirement is domain-specific. With a given confidence threshold for uncertainty control, we can define the corresponding $\sigma$-confident and $\sigma$-uncertain zones. In particular, we first identify the SSC with the largest eigenvalue. It captures the inter-cluster variance and is visualized as the $x$-axis by default. Figure~\ref{fig:sigma} shows that the spatial signatures (top) on this component exhibit a mixture of two normal distributions (bottom). This allows \scheme{} to employ the empirical rule by computing the mean and the standard deviation of each cluster (orange or blue in Figure~\ref{fig:sigma}) and translating the confidence requirement (e.g., $99\%$) into the corresponding intervals (e.g., $3$-$\sigma$ from the mean of the respective cluster). \scheme{} considers the zone between two $\sigma$-confident regions to be uncertain, and any gradient updates from participating clients that fall into this zone will be put on a watchlist. The same client entering the $\sigma$-uncertain zone multiple times will be removed from FL. The same uncertainty management operation can also be done on the temporal signatures to minimize the uncertainty-induced false positives. For clients in the cluster with a smaller dissimilarity score and outside the $\sigma$-uncertain zone, \scheme{} will mark them as malicious and revoke their participation in the remaining rounds.
\begin{table}\small
	\centering
	\renewcommand{\arraystretch}{0.95}
	\setlength\tabcolsep{5.5pt}
	\begin{tabular}{|c|c|c|c|c|}
		\hline
		\multirow{2}{*}{\textbf{\begin{tabular}[c]{@{}c@{}}Perception\\ Poison\end{tabular}}} & \multirow{2}{*}{\textbf{$\boldsymbol{m}$}} & 	\multirow{2}{*}{\textbf{\begin{tabular}[c]{@{}c@{}}Benign\\AP$_\textbf{src}$ (\%)\end{tabular}}} & \multicolumn{2}{c|}{\textbf{Hijacked AP$_\textbf{src}$ (\%)}} \\ \cline{4-5} 
		&  & 	 & 	\textbf{No Defense} & 	\textbf{\scheme{}} \\ \hline
		
		\multirow{3}{*}{\begin{tabular}[c]{@{}c@{}}Class-Poison\\ (VOC)\end{tabular}} & 5\% & 52.43 & 45.35  & 	\textbf{52.41} \\ \cline{2-5} 
		& 10\% & 52.43 & 39.69 & \textbf{52.32} \\ \cline{2-5} 
		& 20\% & 52.43 & 38.22 & \textbf{52.22} \\ \hline
		\multirow{3}{*}{\begin{tabular}[c]{@{}c@{}}BBox-Poison\\ (VOC)\end{tabular}} & 5\%  & 52.43 & 35.76 & \textbf{52.32} \\ \cline{2-5} 
		& 10\% & 52.43 & 23.49 & \textbf{52.15} \\ \cline{2-5} 
		& 20\% & 52.43 & 18.06 & \textbf{52.19} \\ \hline
		\multirow{3}{*}{\begin{tabular}[c]{@{}c@{}}Objn-Poison\\ (VOC)\end{tabular}} & 5\%  & 52.43 & 45.02 & \textbf{51.95} \\ \cline{2-5} 
		& 10\% & 52.43 & 41.61 & \textbf{52.33} \\ \cline{2-5} 
		& 20\% & 52.43 & 37.61 & \textbf{52.29} \\ \hline
		\multirow{3}{*}{\begin{tabular}[c]{@{}c@{}}BBox-Poison\\ (INRIA)\end{tabular}} & 5\%  & 63.49 & 45.03 & \textbf{61.64} \\ \cline{2-5} 
		& 10\% & 63.49 & 32.51 & \textbf{61.25} \\ \cline{2-5} 
		& 20\% & 63.49 & 28.73 & \textbf{61.71} \\ \hline
		\multirow{3}{*}{\begin{tabular}[c]{@{}c@{}}Objn-Poison\\ (INRIA)\end{tabular}} & 5\%  & 63.49 & 61.05 & \textbf{61.98} \\ \cline{2-5} 
		& 10\% & 63.49 & 53.91 & \textbf{62.01} \\ \cline{2-5} 
		& 20\% & 63.49 & 50.18 & \textbf{61.58} \\ \hline
	\end{tabular}\vspace{-0.5em}
	\caption{The AP$_\text{src}$ of FL-trained detectors under the benign scenario (3rd column) and three perception poisoning attacks (4th and 5th columns) with three settings of malicious client percentage $\boldsymbol{m}$.}\label{tab:m}\vspace{-1.5em}
\end{table}

\section{Experimental Evaluation}
We conduct experiments on two object detection benchmark datasets: (i) PASCAL VOC~\cite{everingham2010pascal} and (ii) INRIA Person~\cite{dalal2005histograms}. By default, we have $N=100$ clients and a federated server with $T=200$ rounds for both datasets. At each round, $k=10\%$ of the clients are randomly selected as participants with FedAvg~\cite{mcmahan2017communication} as the aggregation algorithm. Without loss of generality, we focus on Faster R-CNN~\cite{ren2015faster} with VGG16~\cite{simonyan2014very} because perception poisoning is model-agnostic and black-box in nature. For each dataset, we randomly split the training set into $N$ equal-sized partitions and assigned each to one client as its local dataset. We analyze \scheme{} against all three types of perception poisoning on VOC and examine BBox-Poison and Objn-Poison on INRIA as it only supports objects of the ``person" class, which is the default victim source class. The target class in Class-Poison is ``potted plant." For BBox-Poison,  we shrink the bounding box by a factor of $0.10$. By default, $m=20\%$ of the clients in FL are malicious. \scheme{} is operated with a forensic schedule of ten rounds per window, and the confidence required for uncertainty control is set to be $99\%$. 

\subsection{Effectiveness of \scheme{}}
\textbf{Trojan-agnostic Defense.} We first analyze how effective \scheme{} is in mitigating different model hijacking attacks through perception poisoning. {Table~\ref{tab:m}} reports the results measured in average precision~\cite{everingham2010pascal} of the victim source class (AP$_\text{src}$) under three different settings of $m\%$. Given that those attacks can hijack the models to cause a drastic drop in AP$_\text{src}$ ($3$rd and $4$th columns), \scheme{} offers strong defensibility against all of them ($5$th column). 
\begin{figure*}%
	\centering
	\includegraphics[width=0.5\textwidth]{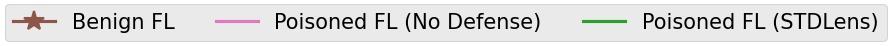}\\
	\begin{subfigure}[b]{0.32\linewidth}
		\centering
		\includegraphics[width=\textwidth]{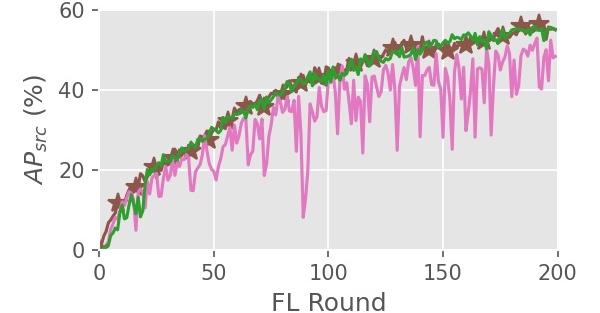}\vspace{0.3em}
		\caption{Class-Poison}
	\end{subfigure}
	\begin{subfigure}[b]{0.32\linewidth}
		\centering
		\includegraphics[width=\textwidth]{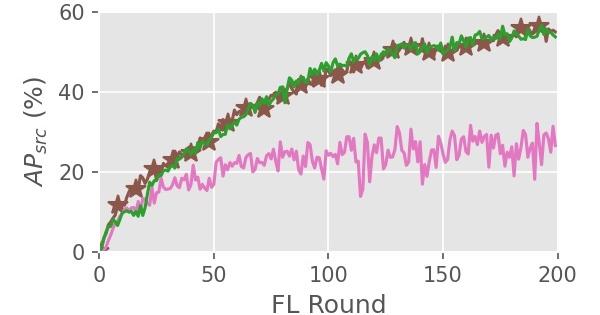}\vspace{0.3em}
		\caption{BBox-Poison}
	\end{subfigure}
	\begin{subfigure}[b]{0.32\linewidth}
		\centering
		\includegraphics[width=\textwidth]{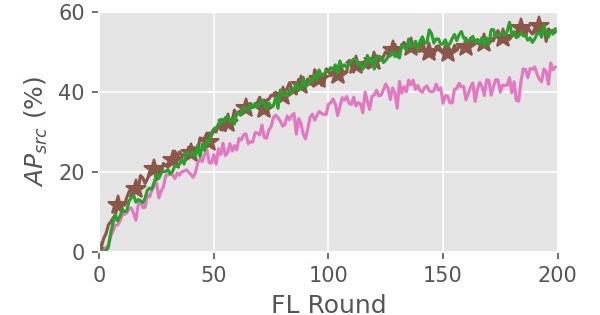}\vspace{0.3em}
		\caption{Objn-Poison}
	\end{subfigure}\vspace{-0.4em}
	\caption{Compared with the benign FL (brown) on VOC, the hijacked FL with no defense (pink) has significant AP$_{\text{src}}$ degradation. \scheme{} (green) purges all malicious clients in an early-round, allows the healing to begin, and reclaims AP$_{\text{src}}$ with a negligible damage.}\vspace{-0.2em}
	\label{fig:exp-general-learning-curve}
\end{figure*}
\begin{table*}\small
\setlength\tabcolsep{1pt}
\renewcommand{\arraystretch}{0.95}
\centering
\begin{tabular}{|cc|cc|cc|}
	\hline
	\multicolumn{2}{|c|}{\textbf{Class-Poison}} & \multicolumn{2}{c|}{\textbf{BBox-Poison}} & \multicolumn{2}{c|}{\textbf{Objn-Poison}} \\ \hline
	\multicolumn{1}{|c|}{\textbf{No Defense}} & \textbf{\scheme{}} & \multicolumn{1}{c|}{\textbf{No Defense}} & \textbf{\scheme{}} & \multicolumn{1}{c|}{\textbf{No Defense}} & \textbf{\scheme{}} \\ \hline
	\multicolumn{1}{|c|}{\includegraphics[width=0.14\linewidth]{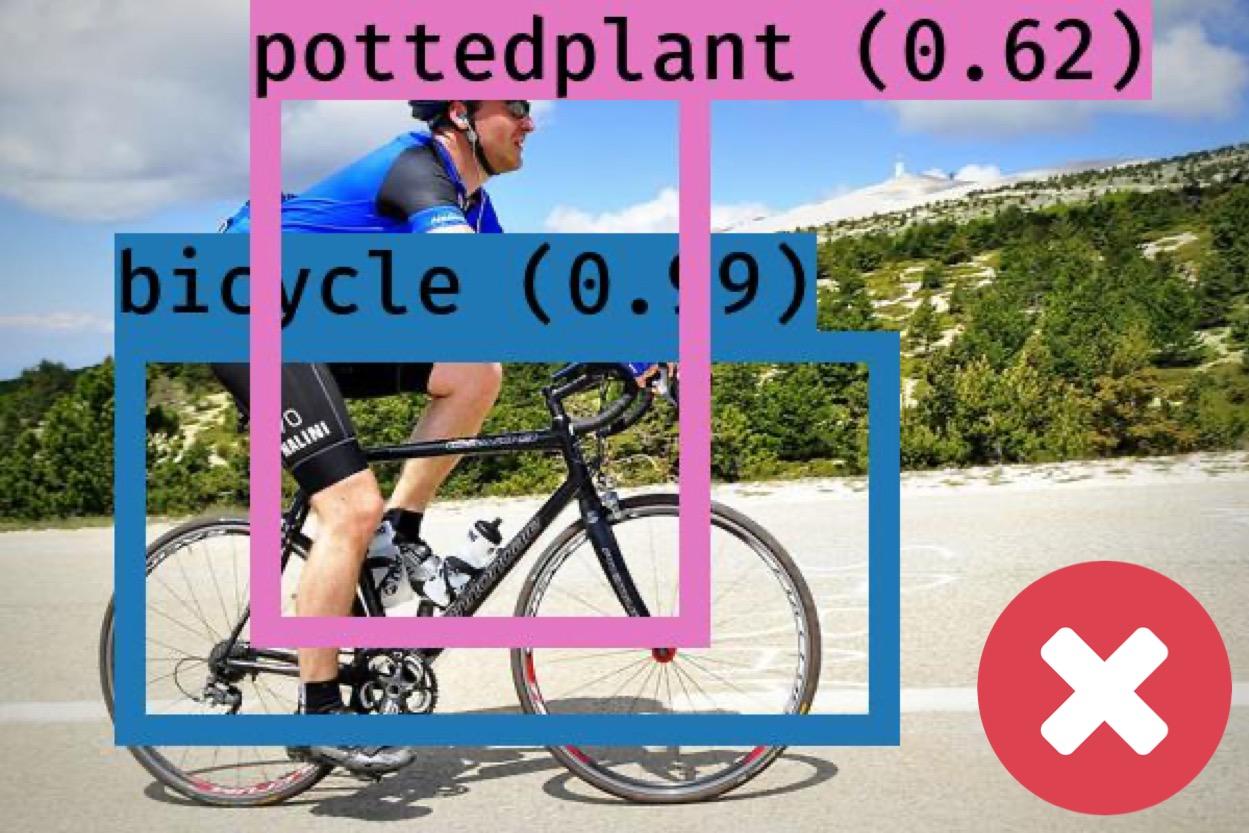}} & \includegraphics[width=0.14\linewidth]{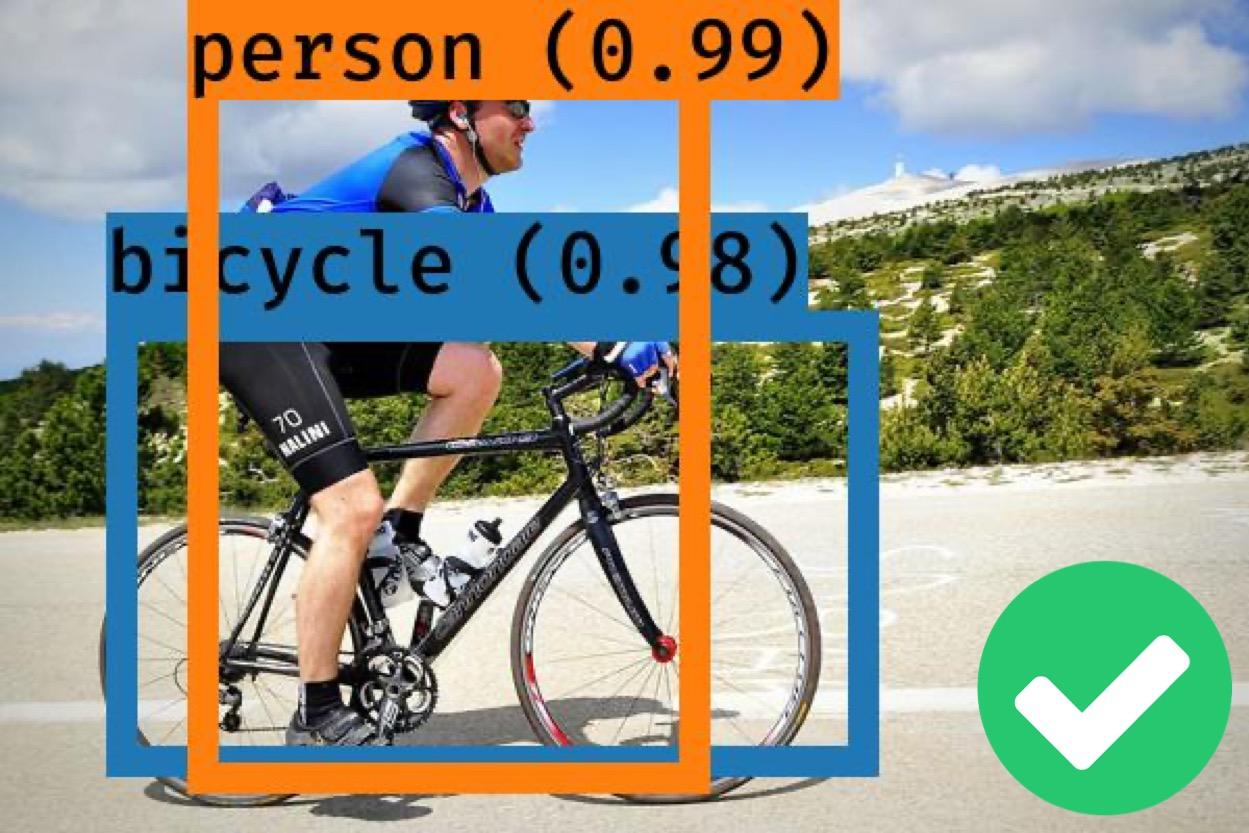} & \multicolumn{1}{c|}{\includegraphics[width=0.14\linewidth]{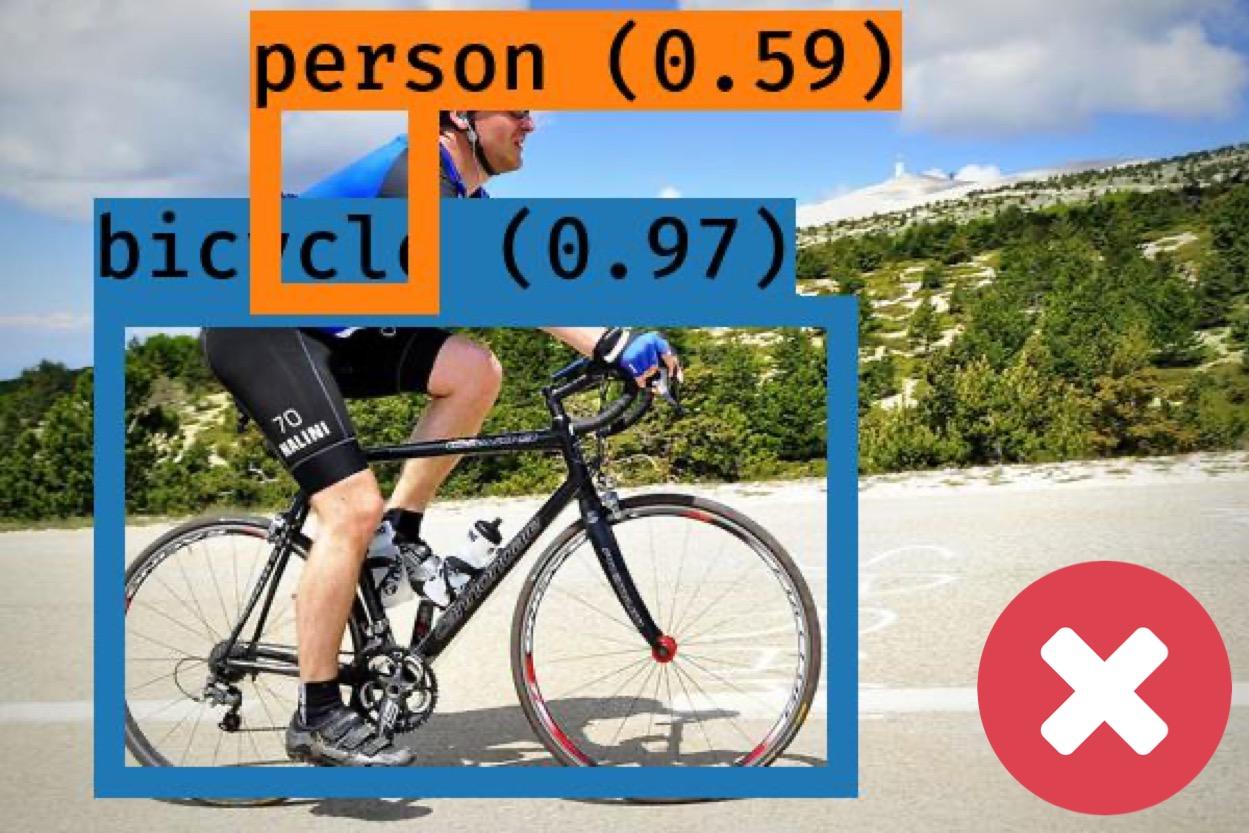}} & \includegraphics[width=0.14\linewidth]{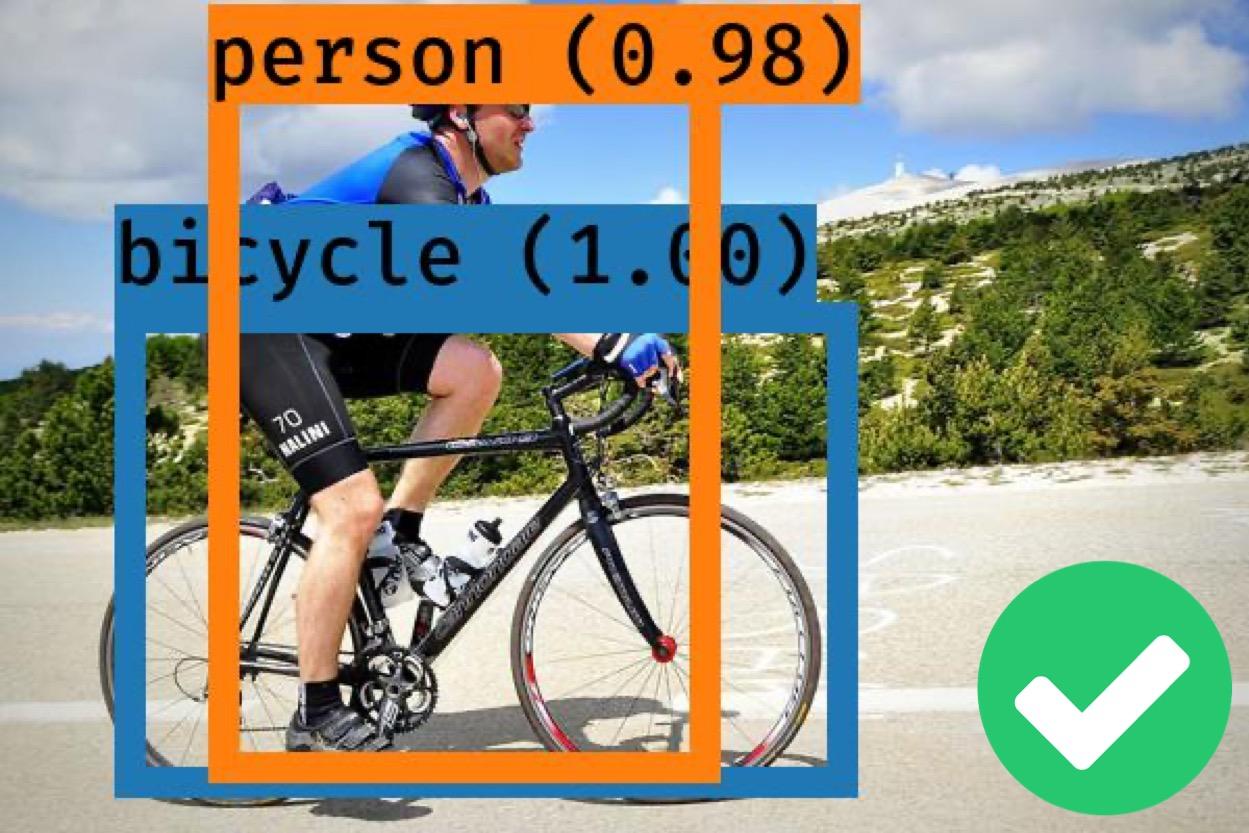} & \multicolumn{1}{c|}{\includegraphics[width=0.14\linewidth]{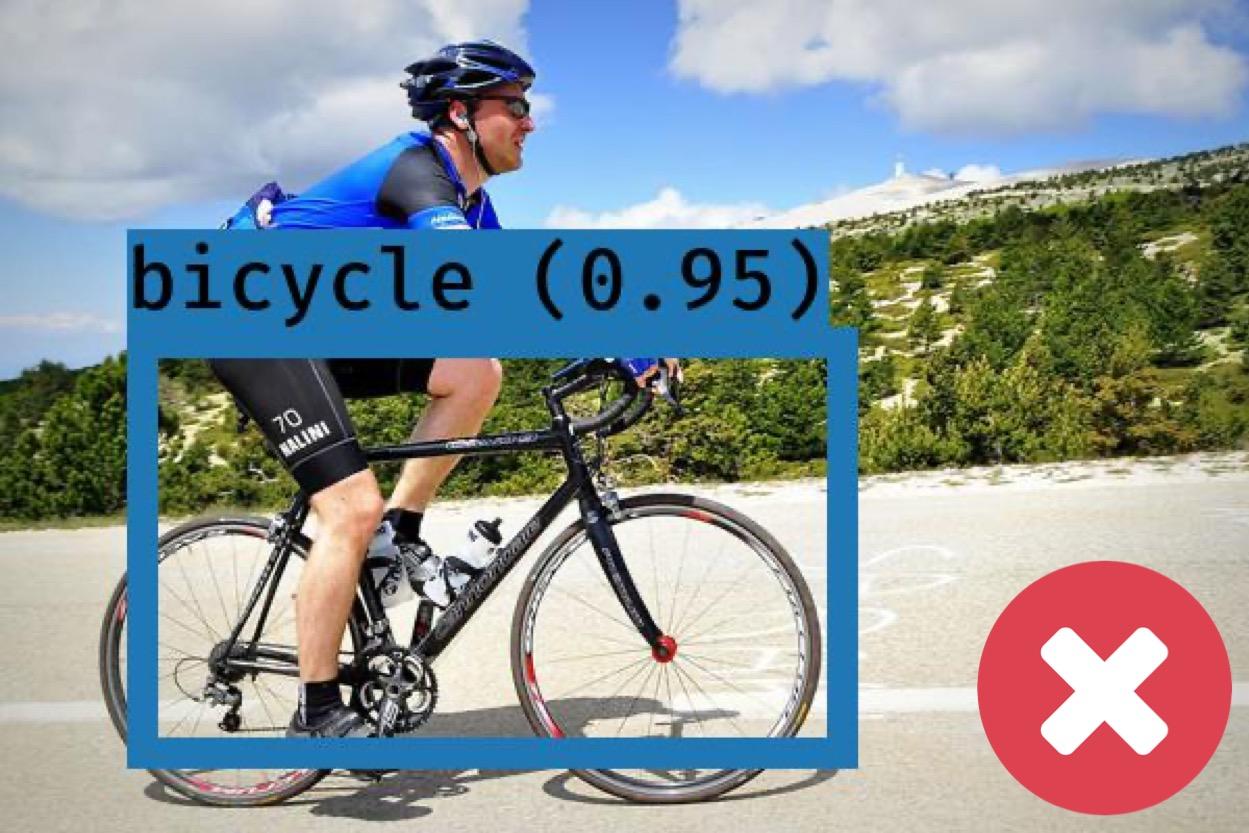}} & \includegraphics[width=0.14\linewidth]{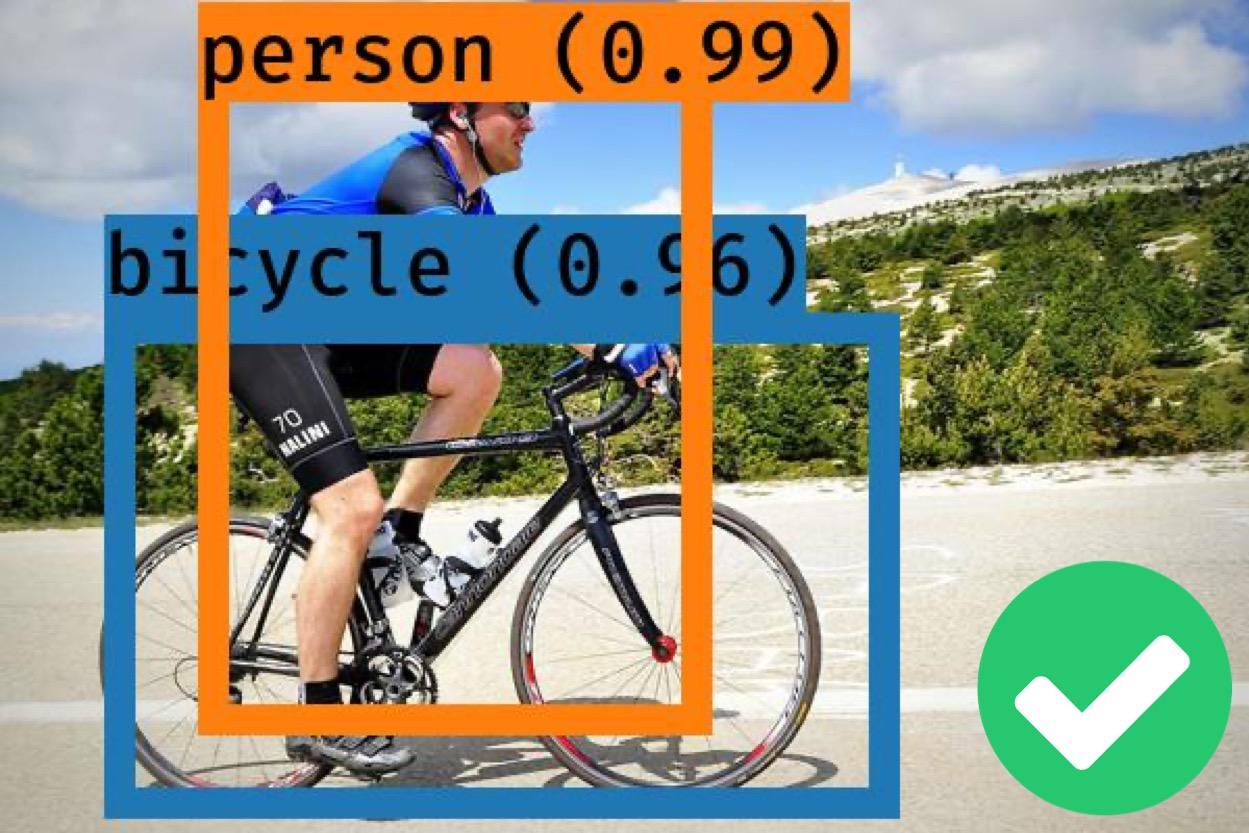} \\ \hline
	\multicolumn{1}{|c|}{\includegraphics[width=0.14\linewidth]{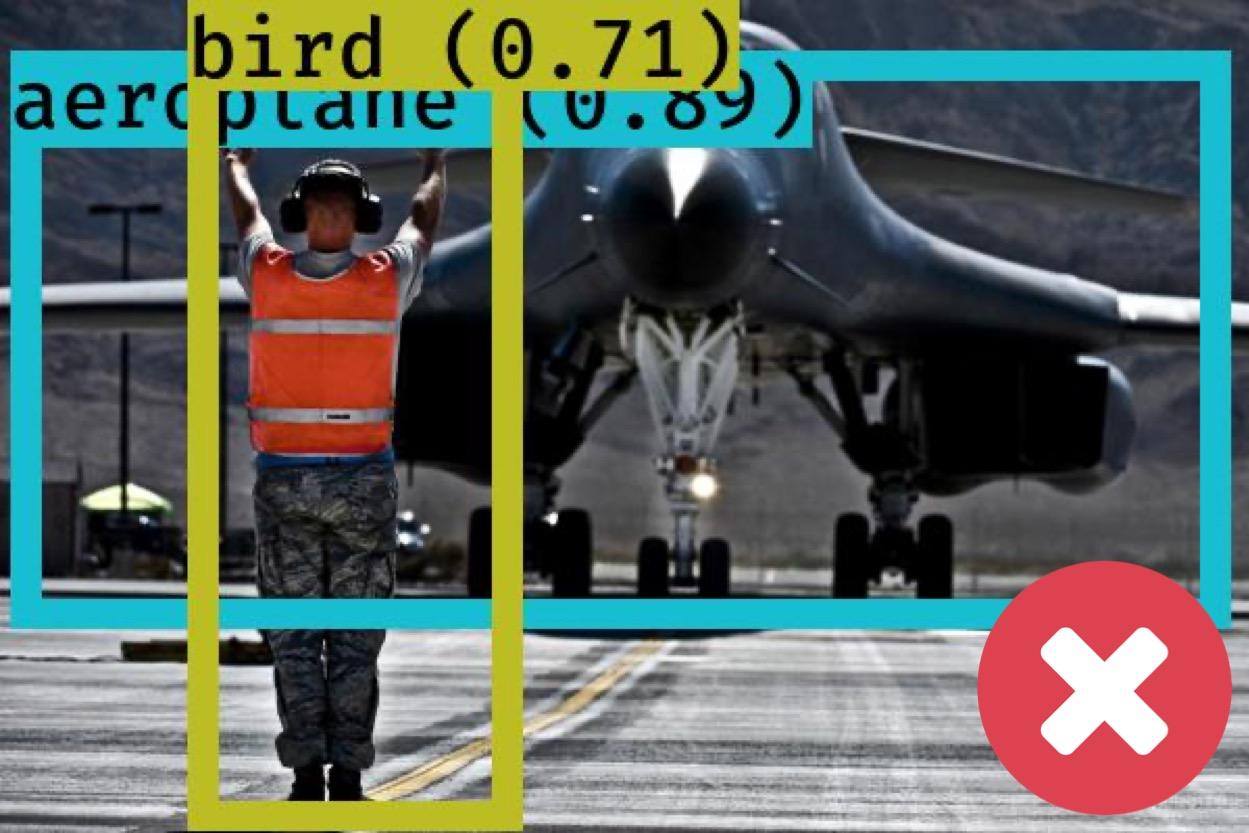}} & \includegraphics[width=0.14\linewidth]{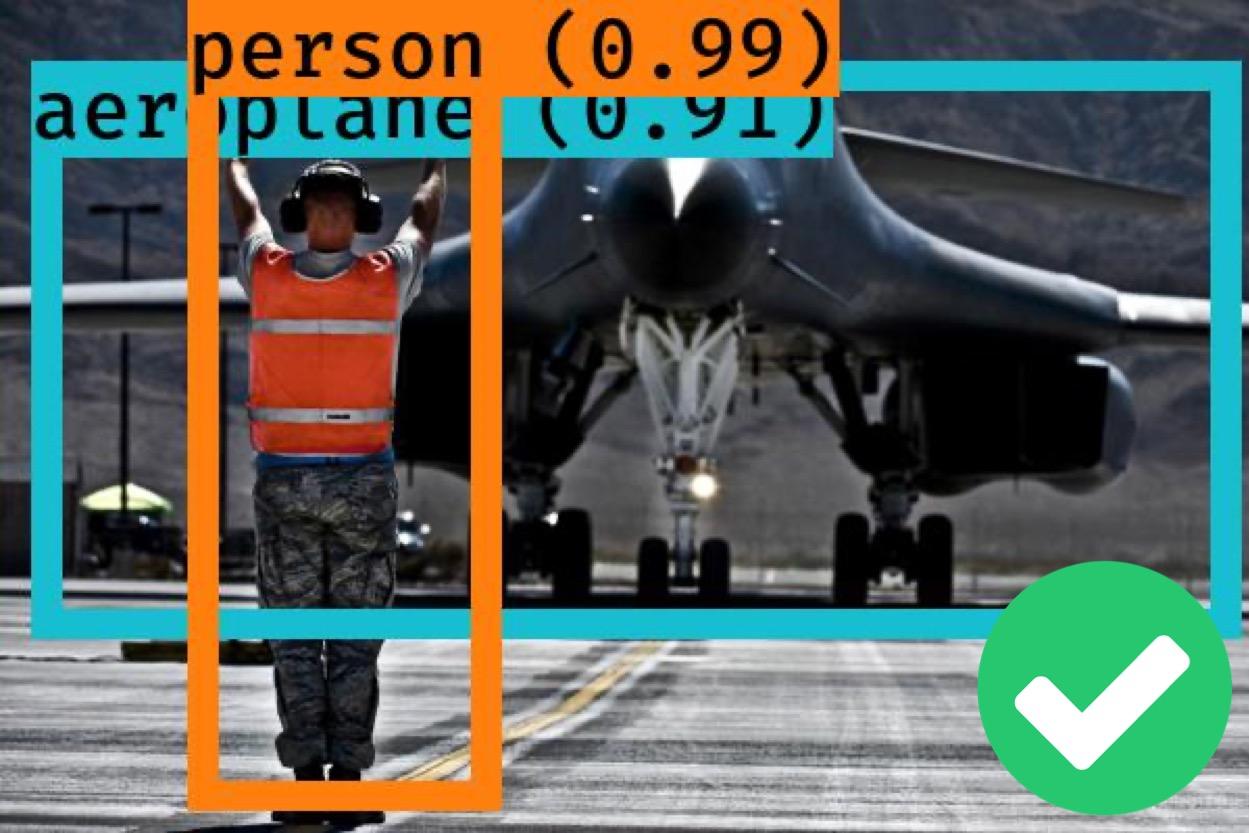} & \multicolumn{1}{c|}{\includegraphics[width=0.14\linewidth]{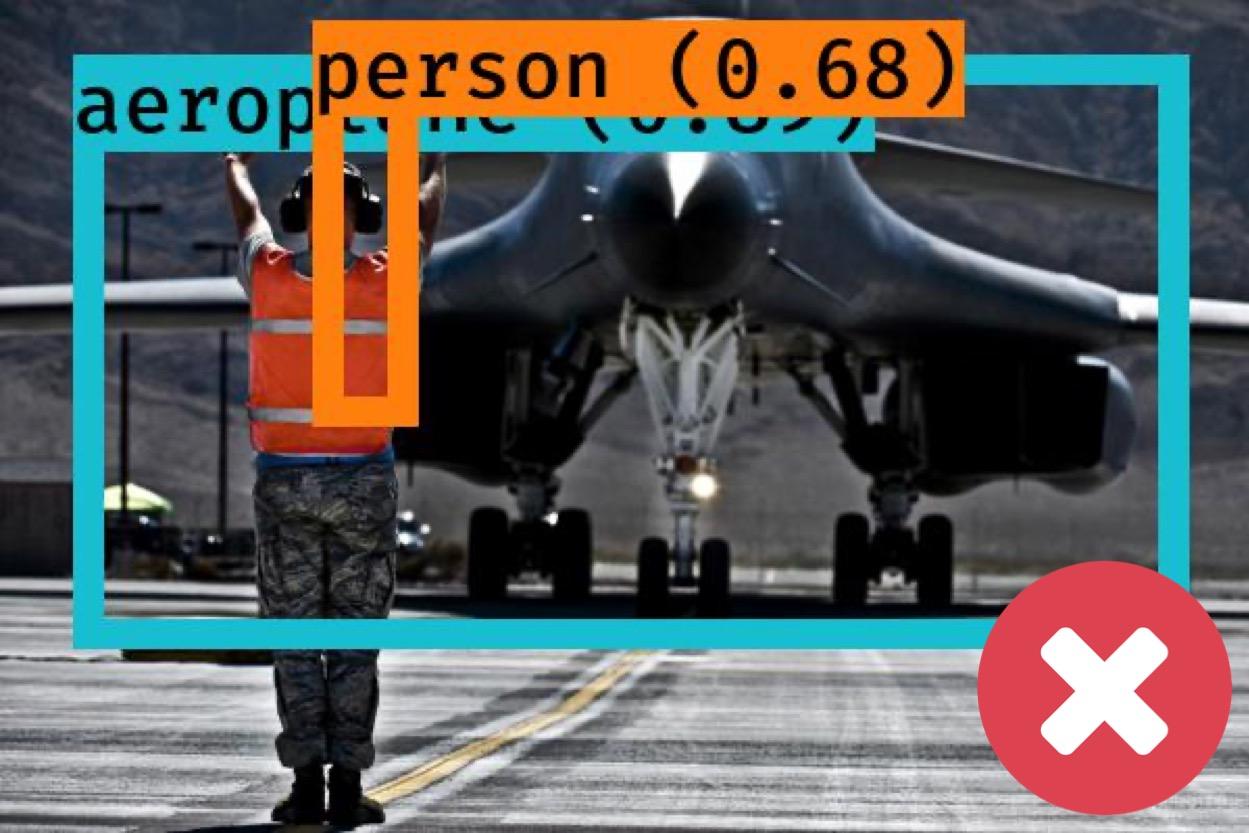}} & \includegraphics[width=0.14\linewidth]{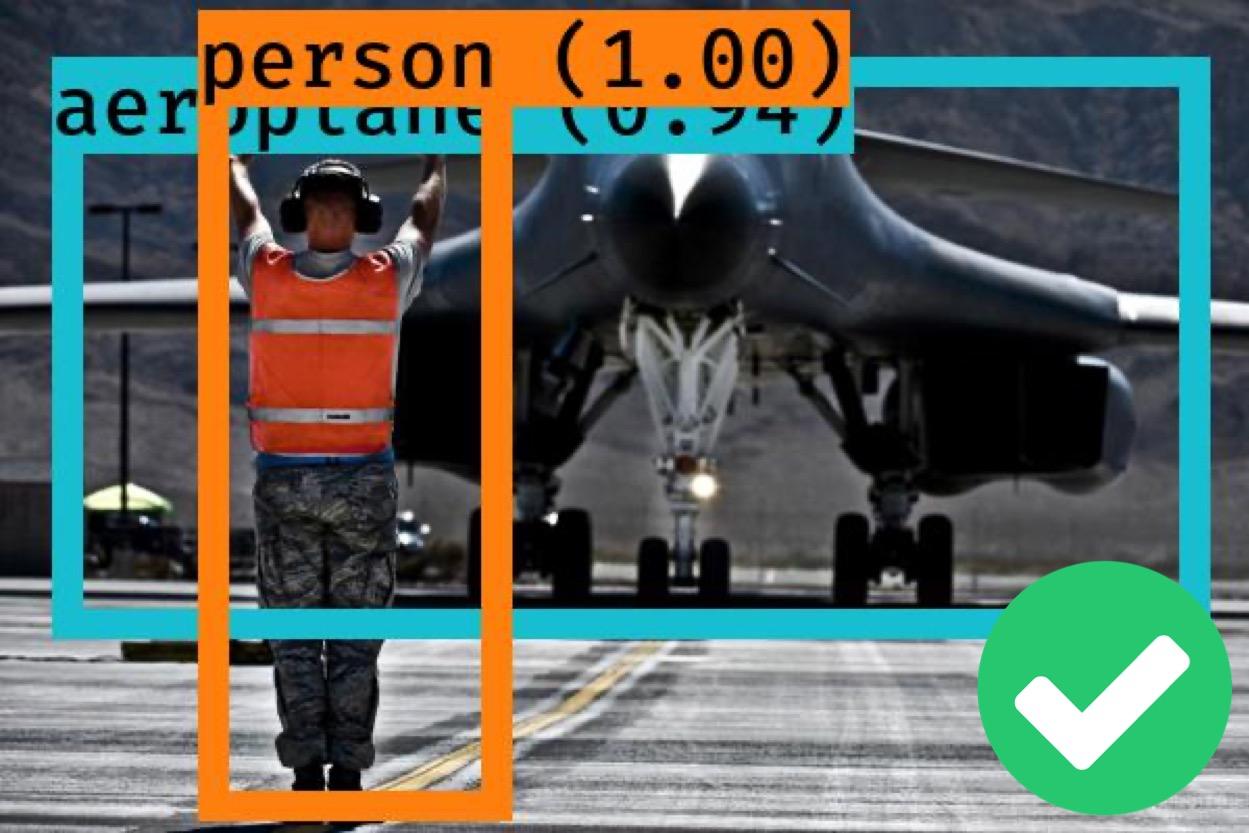} & \multicolumn{1}{c|}{\includegraphics[width=0.14\linewidth]{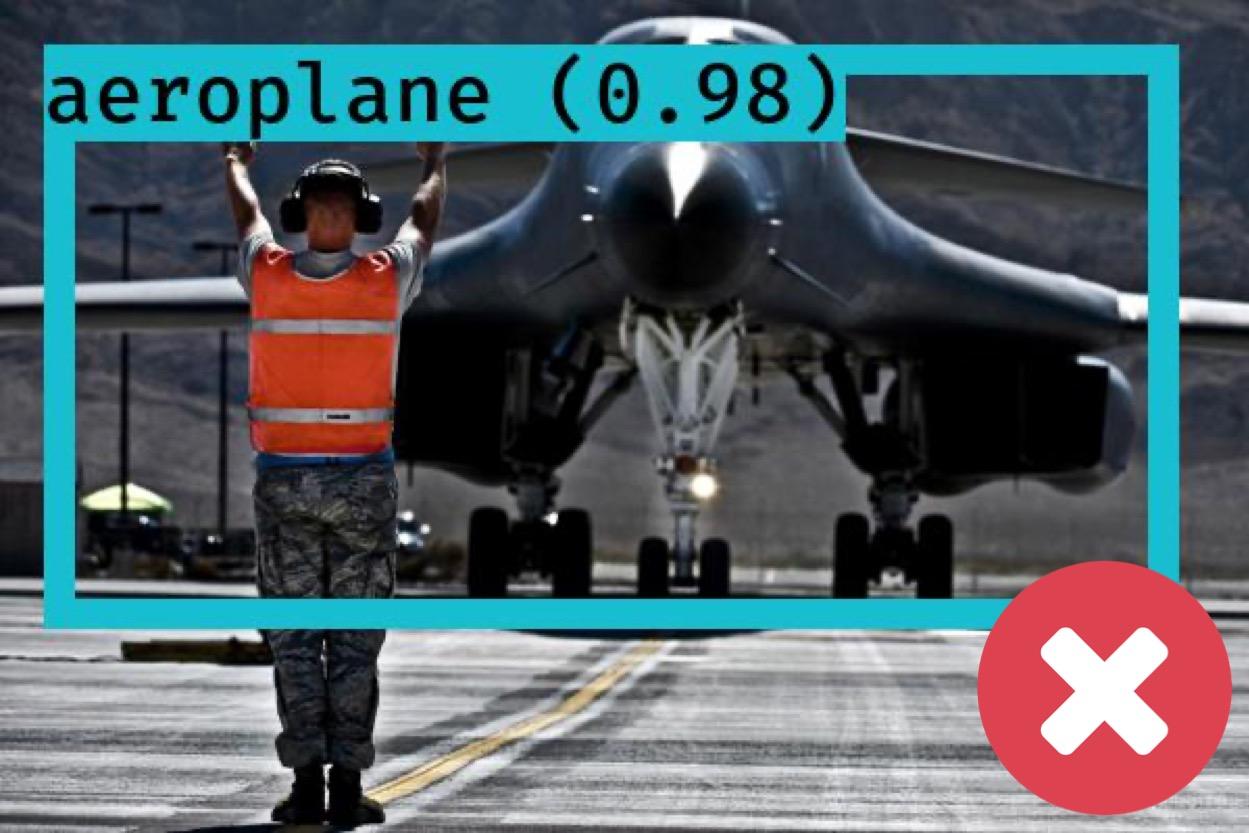}} & \includegraphics[width=0.14\linewidth]{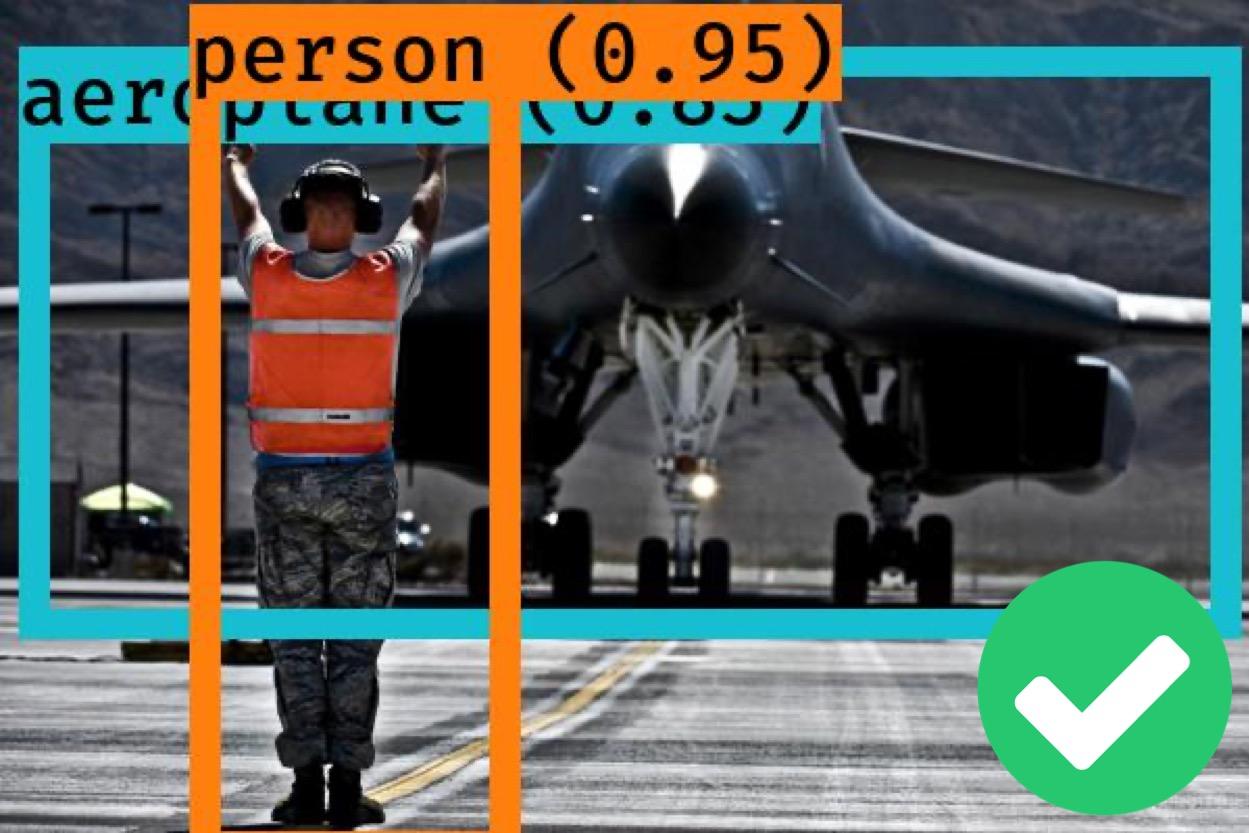} \\ \hline
\end{tabular}\vspace{-0.2em}
\caption{Two images detected by the FL-trained models (VOC) under Class-Poison, BBox-Poison, and Objn-Poison. The FL system with no defense makes designated errors, while the \scheme{}-protected FL preserves correctness even if the adversary attempts to hijack it.}\label{tab:defense-visualization}\vspace{-1em}
\end{table*}
Regardless of which attack is launched and how many malicious clients (reflected by $m$) participate in the attack process, \scheme{} can consistently restore the AP$_\text{src}$, on par with the one obtained in the benign scenario (i.e., $52.43\%$ for VOC and $63.49\%$ for INRIA). For instance, BBox-Poison, which brings down the AP$_\text{src}$ to $18.06\%$ on VOC, is mitigated with an AP$_\text{src}$ of $52.19\%$ under the same attack setting. This validates that \scheme{} is Trojan-agnostic, despite the divergence of different attack effects. Due to similar observations, we focus on VOC in the following discussion.

\textbf{Timely Sanitization and Recovery.} {Figure~\ref{fig:exp-general-learning-curve}} visualizes the learning curves of FL throughout $200$ rounds by comparing three cases: (i) the benign scenario (brown curve), (ii) the hijacked scenario without defense (pink curve), and (iii) the hijacked scenario protected by \scheme{} (green curve). We make the following observations. {First}, Trojaned gradients cause the AP$_\text{src}$ to become either highly unstable (Class-Poison) or consistently degraded (BBox-Poison and Objn-Poison) throughout the FL process. {Second}, when \scheme{} detects all malicious clients at the $30$th round under Class-Poison, the $20$th round under BBox-Poison, and the $30$th round under Objn-Poison, no honest client is misdiagnosed and flagged as malicious. {Third}, the timely sanitization of malicious clients allows the FL system to heal. Even under attacks, the \scheme{}-protected FL can reach an AP$_\text{src}$ comparable with the benign scenario. 

\begin{table*}[]\small
	\centering
	\renewcommand{\arraystretch}{1}
	\begin{tabular}{|c|c|c|c|c|c|c|}
		\hline
		\multirow{2}{*}{\textbf{\begin{tabular}[c]{@{}c@{}}Perception Poison\end{tabular}}} & \multirow{2}{*}{\textbf{\begin{tabular}[c]{@{}c@{}}Object Detection\\with No Defense\end{tabular}}} & \multirow{2}{*}{\textbf{\begin{tabular}[c]{@{}c@{}}Object Detection\\with \scheme{}\end{tabular}}} & \multirow{2}{*}{$\boldsymbol{\beta}$} & \multirow{2}{*}{\textbf{\begin{tabular}[c]{@{}c@{}}Benign\\ AP$_{\textbf{src}}$ (\%)\end{tabular}}} & \multicolumn{2}{c|}{\textbf{Hijacked AP$_{\textbf{src}}$ (\%)}} \\ \cline{6-7} 
		& & & &  & \textbf{No Defense} & \textbf{\scheme{}} \\ \hline
		\multirow{3}{*}{Class-Poison} & \multirow{3}{*}{\includegraphics[width=68.5px]{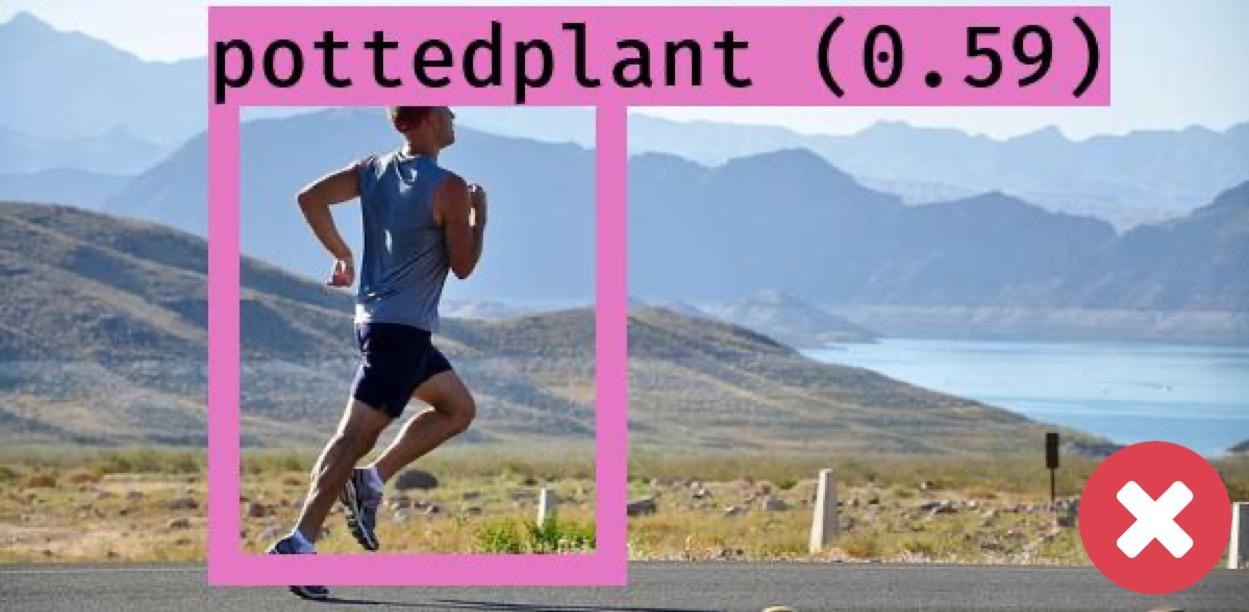}} & \multirow{3}{*}{\includegraphics[width=68.5px]{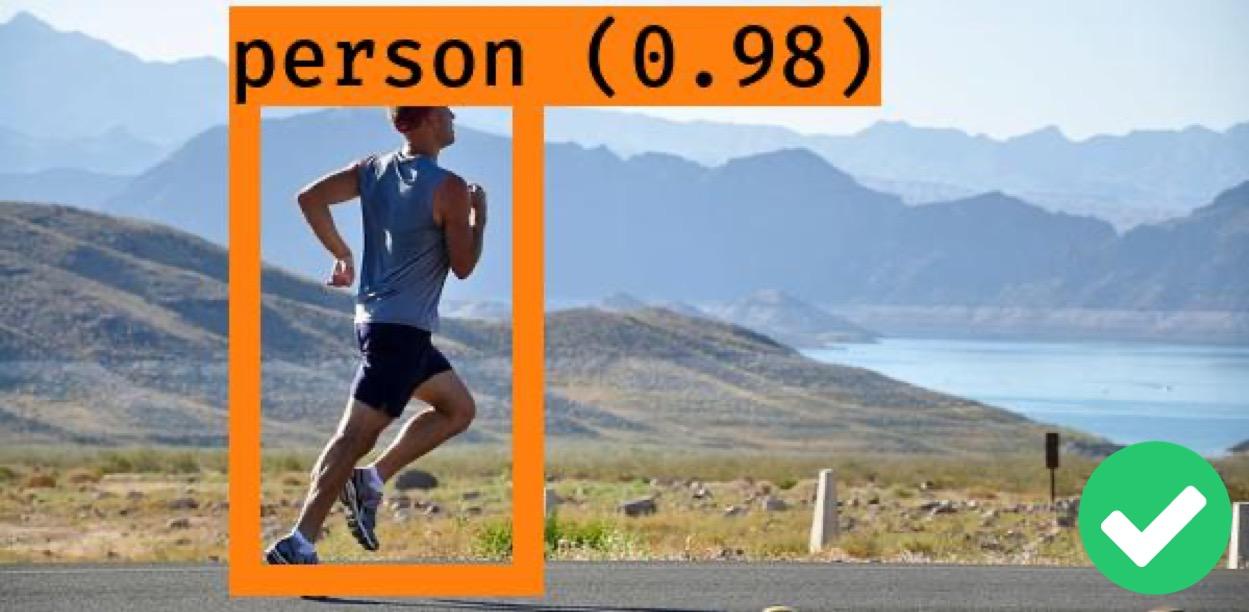}} & \multirow{1}{*}{5\%} & \multirow{1}{*}{52.43} & \multirow{1}{*}{31.83} & \multirow{1}{*}{\textbf{51.74}} \\ \cline{4-7} 
		&  &  & \multirow{1}{*}{10\%} & \multirow{1}{*}{52.43} & \multirow{1}{*}{32.51} & \multirow{1}{*}{\textbf{52.18}} \\  \cline{4-7}  
		&  &  & \multirow{1}{*}{15\%} & \multirow{1}{*}{52.43} & \multirow{1}{*}{34.40} & \multirow{1}{*}{\textbf{51.56}} \\ \hline
		\multirow{3}{*}{BBox-Poison} & \multirow{3}{*}{\includegraphics[width=68.5px]{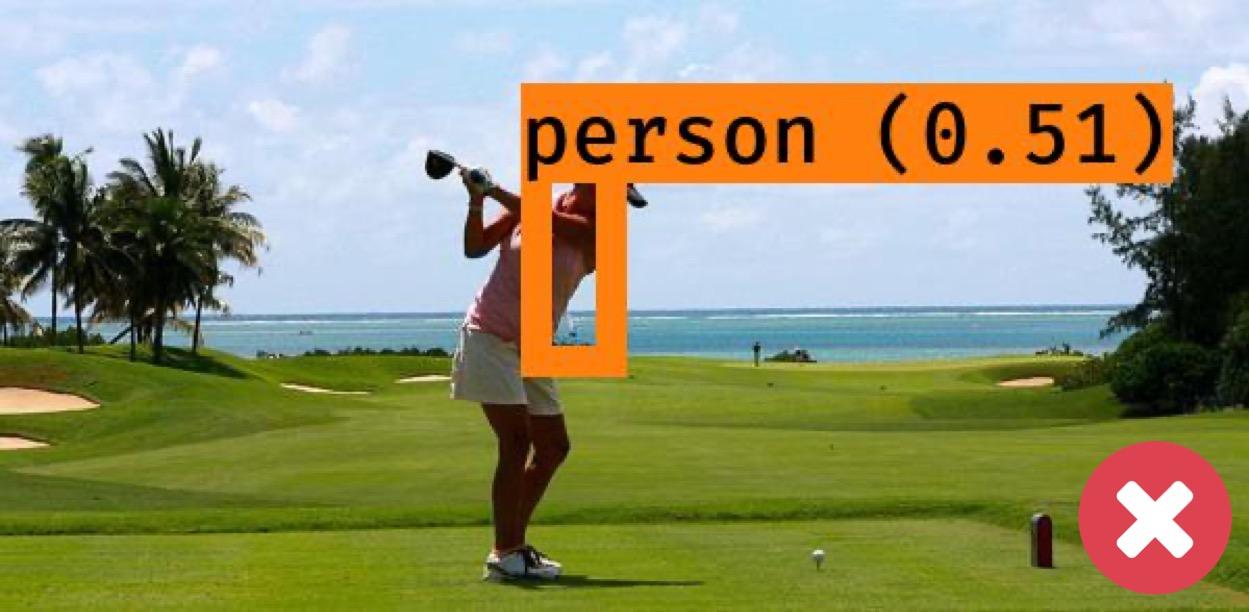}} & \multirow{3}{*}{\includegraphics[width=68.5px]{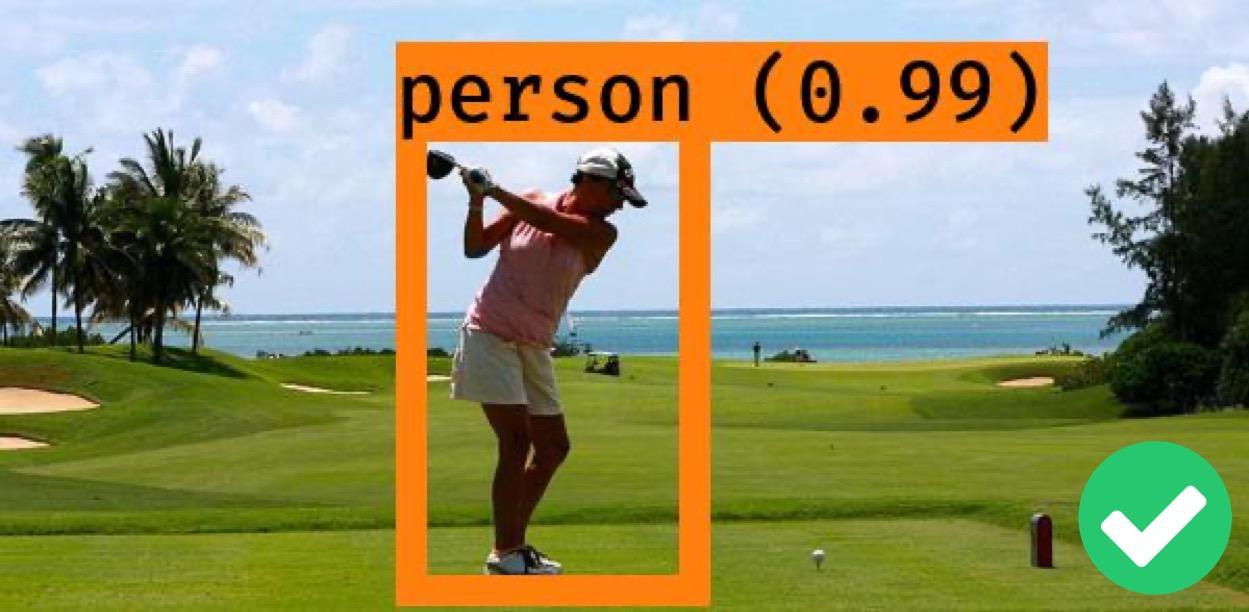}} & \multirow{1}{*}{5\%} & \multirow{1}{*}{52.43} & \multirow{1}{*}{22.36} & \multirow{1}{*}{\textbf{52.39}} \\ \cline{4-7}  
		&  &  & \multirow{1}{*}{10\%} & \multirow{1}{*}{52.43} & \multirow{1}{*}{26.77} & \multirow{1}{*}{\textbf{52.11}} \\  \cline{4-7}  
		&  &  & \multirow{1}{*}{15\%} & \multirow{1}{*}{52.43} & \multirow{1}{*}{27.60} & \multirow{1}{*}{\textbf{51.98}} \\ \hline
		\multirow{3}{*}{Objn-Poison} & \multirow{3}{*}{\includegraphics[width=68.5px]{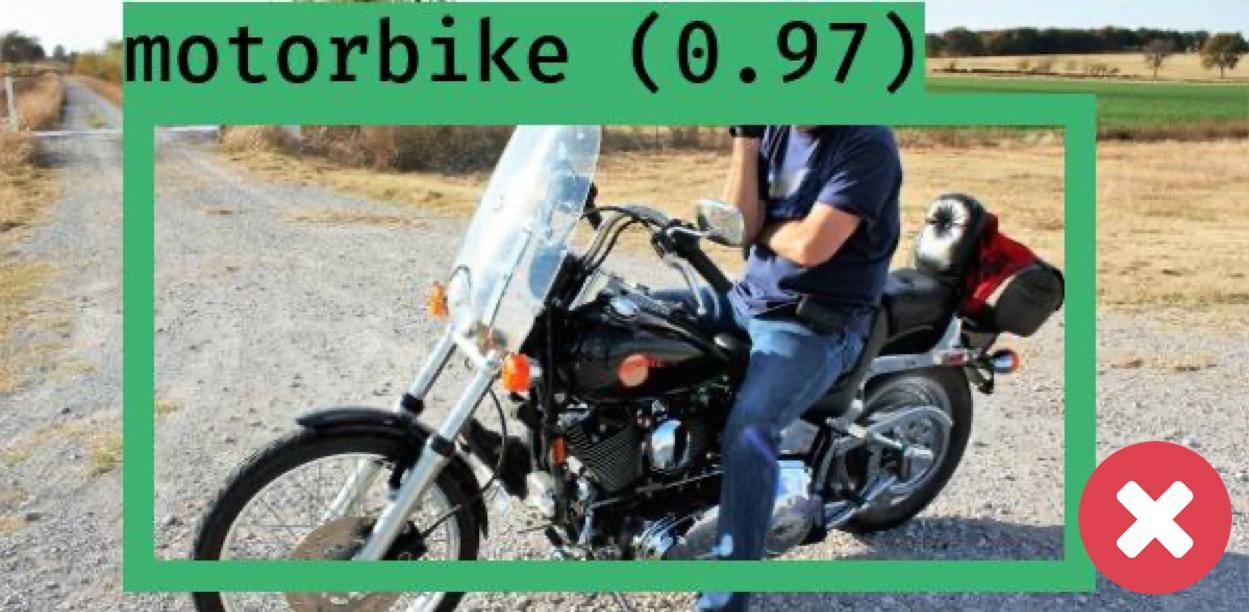}} & \multirow{3}{*}{\includegraphics[width=68.5px]{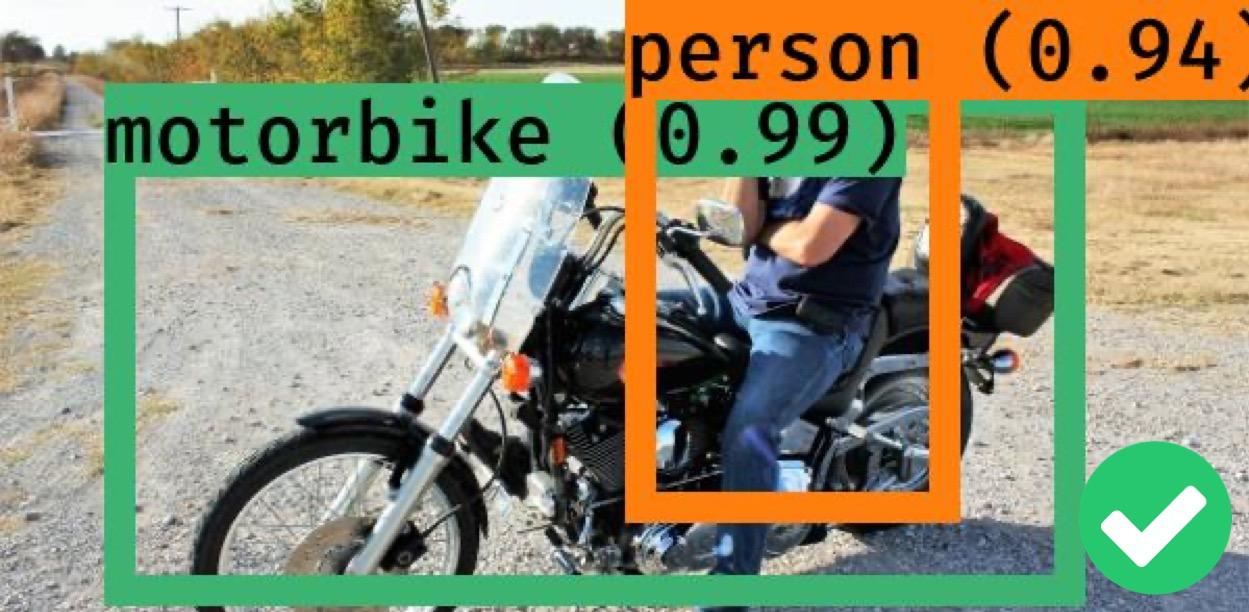}} & \multirow{1}{*}{5\%} & \multirow{1}{*}{52.43} & \multirow{1}{*}{40.96} & \multirow{1}{*}{\textbf{52.05}} \\ \cline{4-7}  
		&  &  & \multirow{1}{*}{10\%} & \multirow{1}{*}{52.43} & \multirow{1}{*}{41.01} & \multirow{1}{*}{\textbf{52.18}} \\ \cline{4-7}  
		&  &  & \multirow{1}{*}{15\%} & \multirow{1}{*}{52.43} & \multirow{1}{*}{41.06} & \multirow{1}{*}{\textbf{52.11}} \\  \hline
	\end{tabular}\vspace{-0.3em}
	\caption{The benign AP$_\text{src}$ and hijacked AP$_\text{src}$ under ${\beta}$-adaptive attacks with varying $\beta$ in different types of perception poison (VOC).}\vspace{-0.8em}
	\label{tab:beta-results}
\end{table*}
\textbf{Visualization.} {Table~\ref{tab:defense-visualization}} visualizes the detection results on two test examples under all three perception poisoning attacks. All three attacks fool the global model with erroneous detection: (i) Class-Poison ($1$st column) fools the model into mislabeling the person as a potted plant (top) or a bird (bottom), (ii) BBox-Poison ($3$rd column) deceives the model to recognize person objects with incorrect bounding boxes, and (iii) Objn-Poison ($5$th column) causes the model to detect no person objects.  In comparison, \scheme{} successfully reinvigorates the FL system with high robustness  ($2$nd, $4$th, and $6$th columns).

\subsection{\scheme{} under Adaptive Attacks}
\begin{figure}
	\centering
	\begin{subfigure}[b]{0.49\linewidth}
		\centering
		\includegraphics[width=\textwidth]{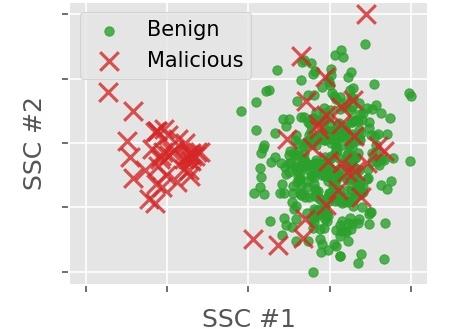}\vspace{0.2em}
		\caption{Spatial Signature Analysis}\label{fig:beta-spatial}
	\end{subfigure}
	\hfill
	\begin{subfigure}[b]{0.49\linewidth}
		\centering
		\includegraphics[width=\textwidth]{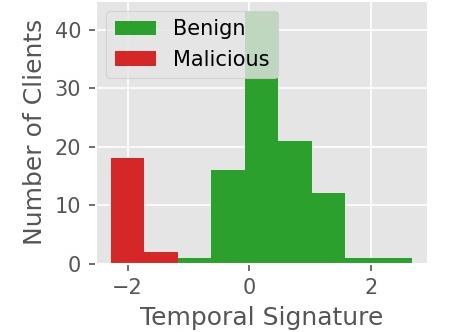}\vspace{0.2em}
		\caption{Temporal Signature Analysis}\label{fig:beta-temporal}
	\end{subfigure}\vspace{-0.4em}
	\caption{$\beta$-adaptive malicious clients randomly camouflage in the benign cluster in the spatial analysis (a), but their statistical characteristics in the temporal signature analysis (b) remain unchanged.}\label{fig:beta}\vspace{-1em}
\end{figure}

\begin{figure}\vspace{-0.5em}
	\begin{subfigure}[b]{0.49\linewidth}
		\centering
		\includegraphics[width=\textwidth]{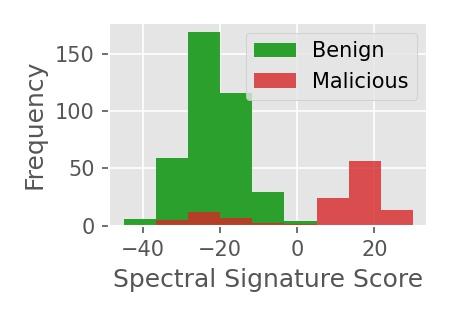}\vspace{-0.2em}
		\caption{$\beta$-adaptive}\label{fig:svd-beta}
	\end{subfigure}
	\hfill
	\begin{subfigure}[b]{0.49\linewidth}
		\centering
		\includegraphics[width=\textwidth]{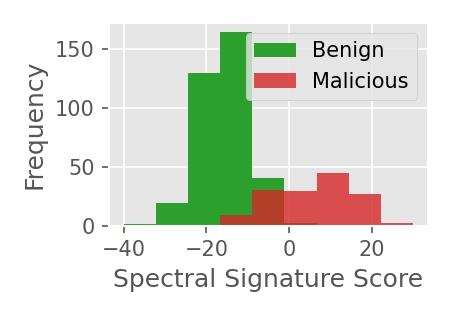}\vspace{-0.2em}
		\caption{$\gamma$-adaptive}\label{fig:svd-gamma}
	\end{subfigure}\vspace{-.4em}
	\caption{Spectral signatures~\cite{tran2018spectral} fail under (a) $\beta$-adaptive and (b) $\gamma$-adaptive attacks as the statistical properties of malicious contributions (red) overlap with benign ones (green).}\label{fig:svd}\vspace{-0.5em}
\end{figure}

\vspace{-0.5em}\textbf{$\boldsymbol{\beta}$-adaptive Attacks.} An immediate question we need to address is the defensibility of \scheme{} against adaptive adversaries who strategically change their attack methods.  For instance, malicious clients could camouflage in the crowd of benign clients by occasionally contributing non-Trojaned gradients. We refer to this type of adaptive adversary as ${\beta}$-adaptive. When a $\beta$-adaptive adversary participates in an FL round, it has, e.g., a $\beta=10\%$ chance of not poisoning the local dataset during the local model training. As a result, this adaptive adversary will contribute to the cluster with benign gradient updates and also to the cluster with Trojaned gradient updates, aiming to fool the spatial and temporal signature analysis. Indeed, such an adaptive attack strategy can jeopardize the spatial analysis because malicious clients can be mistakenly clustered with honest ones, as depicted in {Figure~\ref{fig:beta-spatial}}.
This makes defenses using solely spatial clustering~\cite{shen2016auror,tolpegin2020data} unable to purge all malicious clients. Even for another representative defense using spectral signatures~\cite{tran2018spectral}, the statistical properties of malicious contributions become the same as those benign ones, as shown in {Figure~\ref{fig:svd-beta}}. However, the temporal signature analysis in \scheme{} is, by design, resilient to such adaptive strategies, as shown in {Figure~\ref{fig:beta-temporal}}. Recall in Equation~\ref{eq:td} that the temporal signature of a client is computed w.r.t. each cluster, and the one with a smaller magnitude is taken to be the lower bound of malicious likelihood. Hence, the statistical characteristics of honest and malicious clients can be preserved in our temporal analysis, where malicious clients (red) have a much smaller temporal signature than honest clients (green). {Table~\ref{tab:beta-results}} compares the hijacked AP$_\text{src}$ with the $\beta$-adaptive attack under no defense scenario ($6$th column) with the FL system under the protection by \scheme{} ($7$th column). A total of nine scenarios under three attacks are compared by varying the configurations of $\beta$ from $5\%$ to $15\%$. We make two observations. {First}, under the $\beta$-adaptive attack, the AP$_\text{src}$ suffers from significant degradation, from $52.43\%$ to $22.36\%$, even though malicious clients do not send Trojaned gradients all the time. For adaptive adversaries, the choice of $\beta$ should be kept small as a larger $\beta$ can notably weaken the attack effectiveness. {Second}, \scheme{} remains resilient under $\beta$-adaptive adversaries since it can consistently restore the AP$_\text{src}$ to at least $51.56\%$ for all nine scenarios, on par with that under non-adaptive attacks (recall Table~\ref{tab:m}). This is primarily because our temporal analysis is robust against such a camouflage. Integrating spatio-temporal statistics allows the identification of the malicious cluster to remain accurate under $\beta$-adaptive attacks. 

\textbf{$\boldsymbol{\gamma}$-adaptive Attacks.} Another attempt to launch adaptive attacks is to wash out the spatial clustering characteristics by only poisoning $\gamma\%$ of training samples in the local dataset of a malicious client. Under this adaptive setting, the contribution from a malicious client becomes the gradient updates computed over a mixture of clean and poisoned training samples. We conduct experiments with two settings of $\gamma$ with results in {Table~\ref{tab:gamma}}. 
\begin{table}\small
\centering
\renewcommand{\arraystretch}{0.95}
\begin{tabular}{|c|c|c|c|}
	\hline
	\multirow{2}{*}{\textbf{$\boldsymbol{\gamma}$}} & \multirow{2}{*}{\textbf{\begin{tabular}[c]{@{}c@{}}Benign\\ AP$_{\textbf{src}}$ (\%)\end{tabular}}} & \multicolumn{2}{c|}{\textbf{Hijacked AP$_{\textbf{src}}$ (\%)}} \\ \cline{3-4} 
	& & \textbf{No Defense} & \textbf{\scheme{}} \\ \hline
	80\% & 52.43 & 40.03 & \textbf{52.09} \\ \hline
	60\% & 52.43 & 46.01 & \textbf{51.54} \\ \hline
\end{tabular}\vspace{-0.6em}
\caption{Quantitative studies of $\boldsymbol{\gamma}$-adaptive Class-Poison (VOC).}\label{tab:gamma}\vspace{-1.6em}
\end{table}
The {${\gamma}$-adaptive attacks can bring down the AP$_\text{src}$ from $52.43\%$ to $40.03\%$ and $46.01\%$ when we set {${\gamma}$ to $80\%$ and $60\%$, respectively. More malicious clients fall inside the $\sigma$-uncertain zone, which will be incorrectly identified by defenses using only spatial clustering analysis~\cite{shen2016auror,tolpegin2020data}. Such an adaptive attack also makes spectral signatures~\cite{tran2018spectral} ambiguous, as the statistical properties of benign and malicious gradients overlap (Figure~\ref{fig:svd-gamma}). For those uncertain clients, \scheme{} has to put the decision on hold and wait for more rounds to accumulate additional observations for making confident decisions. Table~\ref{tab:gamma} shows that \scheme{} remains robust under  {${\gamma}$-adaptive attacks because all malicious clients can still be identified with no false alarm. The AP$_\text{src}$ can be recovered to reach over $51.54\%$, comparable to the benign AP$_\text{src}$ of $52.43\%$.
		
\textbf{Late-stage Adaptive Attacks.} Instead of starting the poisoning process from the beginning of FL, an adaptive adversary could only start poisoning the training data when selected in the later rounds of FL. In this case, the FL system has no Trojaned gradient updates contributed by its clients for a period of time. We conduct such a late-stage adaptive attack by running Class-Poison after the $100$th round during the FL process. {Figure~\ref{fig:late-round-learning-curve}} (left) shows the learning curves of AP$_\text{src}$ for three scenarios: (i) the benign scenario, (ii) the late-stage adaptive attack without defense, and (iii) the one protected by \scheme{}. Considering the no-defense scenario (pink curve), the AP$_\text{src}$ fluctuates wildly after the $100$th round. In comparison, \scheme{} (green curve) starts detecting the existence of Trojans from the window of rounds 101-110 (top right) and removes the identified malicious clients. It successfully expels all anomalies at the $130$th round and allows the FL to heal as no more malicious clients attempt to mislead the FL system (e.g., the window of rounds 131-140 at the bottom right).
\begin{figure}
	\centering
	\includegraphics[width=\linewidth]{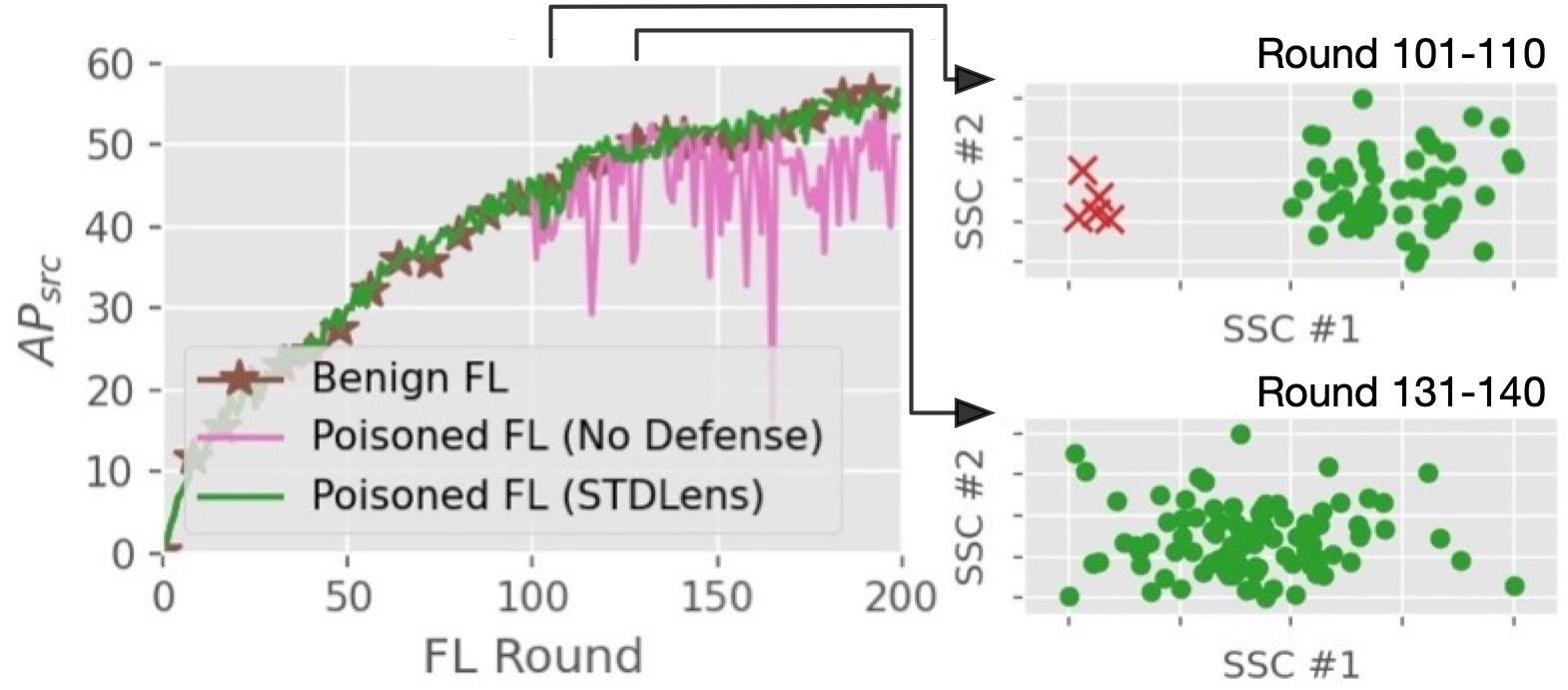}\vspace{-0.7em}
	\caption{The malicious clients become active after the $100$th round (VOC), causing the AP$_{\text{src}}$ to drop. \scheme{} purges all malicious clients at the $130$th round and allows the FL to recover.}\label{fig:late-round-learning-curve}\vspace{-1em}
\end{figure}

\subsection{Comparisons with Existing Defenses}
We further compare \scheme{} with two existing defense methods: spatial clustering~\cite{tolpegin2020data} and spectral signature~\cite{tran2018spectral}, which can be extended to FL-based object detection. {Table~\ref{tab:baseline}} reports the quantitative results. All three defense solutions can detect all malicious clients over the course of FL (i.e., achieving the defense recall of $1.00$). \scheme{} completes the detection and revocation of malicious clients between $20$ to $30$ rounds, and spatial clustering takes $30$ to $40$ rounds to detect and remove anomalies, while malicious clients can stay alive in the FL system for at most $80$ rounds with spectral signature. The main differentiation factor lies in defense precision. Even though \scheme{} takes a slightly longer execution time (i.e., $452$ms with $84$ms for temporal signature analysis and $44$ms for $\sigma$-density inspection), it successfully identifies all malicious clients without mistakenly rejecting any benign clients (false positives), achieving a defense precision of $1.00$. This allows the \scheme{}-protected FL to train a global object detector with high accuracy, as shown in the $2$nd column in {Table~\ref{tab:comp-vis}}. In comparison, the spatial clustering approach can reach a low precision of $0.53$ under Objn-Poison, due to gradient contributions falling in the zone between the two clusters, causing ambiguity in anomaly detection. This result shows that defenses based solely on spatial clustering may not offer stable robustness. The spectral signature defense performs the worst, with a defense precision ranging from $0.26$ to $0.54$. It implies that while the defense can expel malicious clients, a considerable number of honest ones are also mistakenly rejected. One reason is the changing malicious population over time due to the removal of malicious clients in the early rounds. The excessive removal of clients can weaken the benign learning signal and degrade the FL-trained model, as shown in the $3$rd and $4$th columns in Table~\ref{tab:comp-vis}. 
\begin{table}[]\small
	\centering
	\setlength\tabcolsep{4pt}
	\renewcommand{\arraystretch}{0.95}
	\begin{tabular}{|l|c|c|c|}
		\hline
		\multicolumn{1}{|c|}{\textbf{}} & \multicolumn{3}{c|}{\textbf{Defense Precision@Round w/ Max. Recall}} \\ \cline{2-4} 
		\multicolumn{1}{|c|}{\textbf{}} & \textbf{\begin{tabular}[c]{@{}c@{}}\scheme{}\\ (Ours)\end{tabular}} & \textbf{\begin{tabular}[c]{@{}c@{}}Spatial\\Clustering~\cite{tolpegin2020data}\end{tabular}} & \textbf{\begin{tabular}[c]{@{}c@{}}Spectral\\Signature~\cite{tran2018spectral}\end{tabular}} \\ \hline
		Class-Poison & \textbf{1.00@30} & 0.95@30 & 0.53@40 \\ \hline
		BBox-Poison & \textbf{1.00@20} & 0.71@30 & 0.26@80 \\ \hline
		Objn-Poison & \textbf{1.00@30} & 0.53@40 & 0.54@40 \\ \hhline{=|=|=|=}
		\textbf{Time (ms)} & 452 & 342 & 1423 \\ \hline
	\end{tabular}\vspace{-0.5em}
	\caption{Comparing \scheme{} with two state-of-the-art methods on VOC in terms of the defense precision at the round where the maximum recall is achieved and the execution time (the last row).}\label{tab:baseline}\vspace{-0.5em}
\end{table}

\begin{table}[]\small
	\centering
	\setlength\tabcolsep{2pt}
	\renewcommand{\arraystretch}{0.95}
	\begin{tabular}{|l|ccc|}
		\hline
		\multirow{2}{*}{} & \multicolumn{3}{c|}{\textbf{Object Detection Results by Hijacked Models}}                                                                                                                                                                                                                                                                               \\ \cline{2-4} 
		& \multicolumn{1}{c|}{\textbf{\begin{tabular}[c]{@{}c@{}}STDLens\\(Ours)\end{tabular}}} & \multicolumn{1}{c|}{\textbf{\begin{tabular}[c]{@{}c@{}}Spatial\\ Clustering~\cite{tolpegin2020data}\end{tabular}}} & \multicolumn{1}{c|}{\textbf{\begin{tabular}[c]{@{}c@{}}Spectral\\ Signature~\cite{tran2018spectral}\end{tabular}}} \\ \hline
		\begin{tabular}[c]{@{}l@{}}Class-\\Poison\end{tabular}      &  \multicolumn{1}{c|}{\raisebox{-.4\height}{\includegraphics[width=64px]{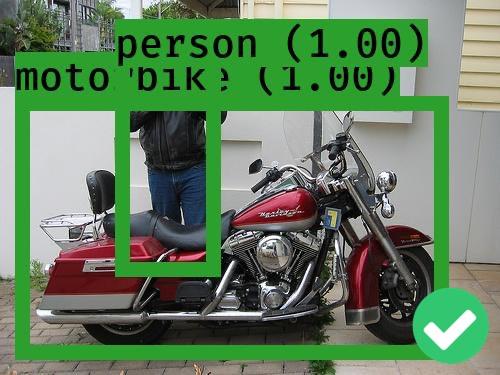}}}& \multicolumn{1}{c|}{\raisebox{-.4\height}{\includegraphics[width=64px]{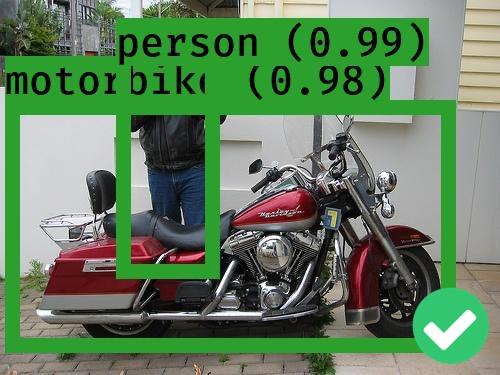}}}&  \raisebox{-.4\height}{\includegraphics[width=64px]{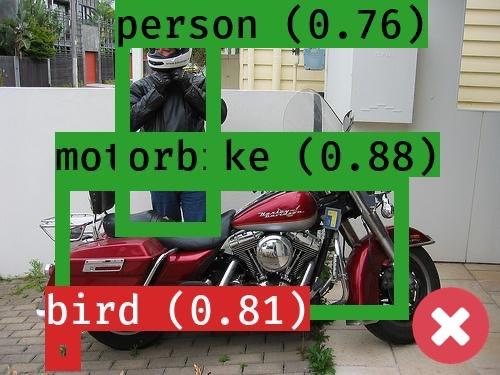} }                                                                                          \\ \hline
		\begin{tabular}[c]{@{}l@{}}BBox-\\Poison\end{tabular}       &  \multicolumn{1}{c|}{\raisebox{-.4\height}{\includegraphics[width=64px]{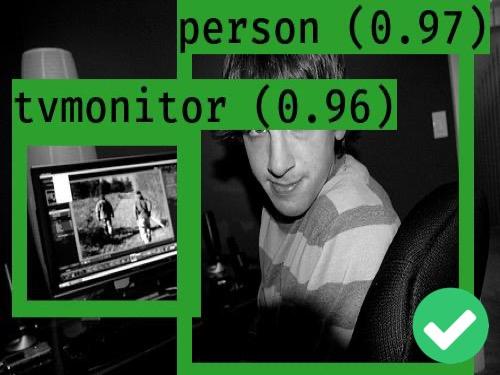}}} & \multicolumn{1}{c|}{\raisebox{-.4\height}{\includegraphics[width=64px]{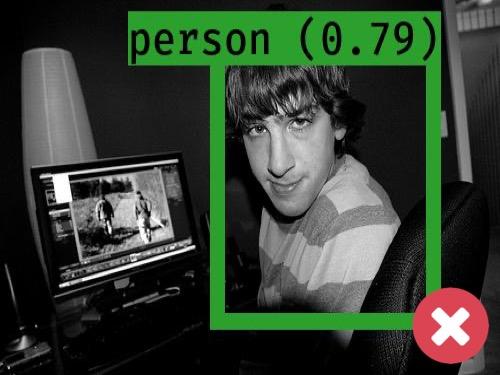}}} &  \raisebox{-.4\height}{\includegraphics[width=64px]{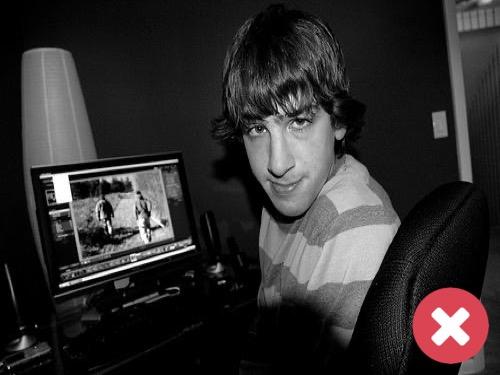} }                                                                                          \\ \hline
		\begin{tabular}[c]{@{}l@{}}Objn-\\Poison\end{tabular}       &  \multicolumn{1}{c|}{\raisebox{-.4\height}{\includegraphics[width=64px]{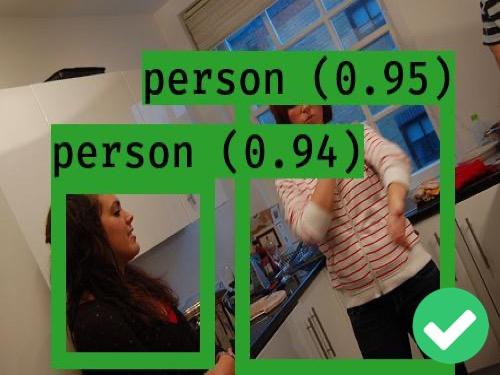}}} & \multicolumn{1}{c|}{\raisebox{-.4\height}{\includegraphics[width=64px]{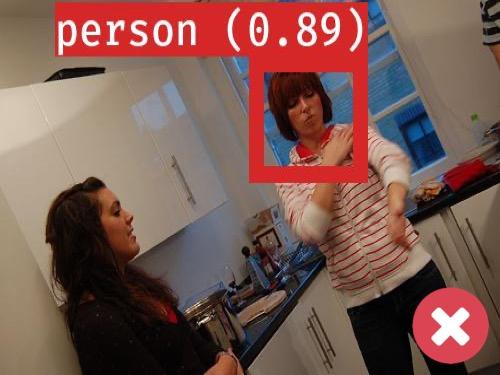}}} &       \raisebox{-.4\height}{\includegraphics[width=64px]{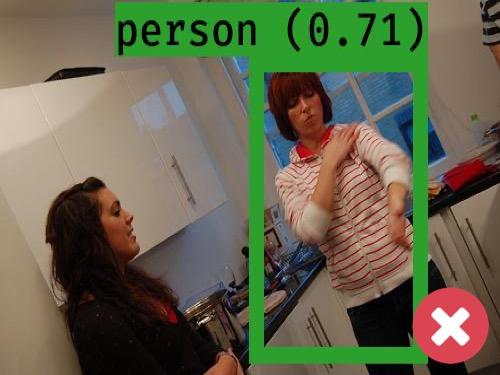}}                                                                                      \\ \hline
	\end{tabular}\vspace{-0.5em}
	\caption{The high defense precision offered by \scheme{} allows the FL-trained model to remain accurate. The excessive removal of clients by  other defenses incurs performance degradation (VOC).}\vspace{-1.5em}\label{tab:comp-vis}
\end{table}

\section{Conclusions}
With the goal of building a trustworthy federated training process, we have presented \scheme{}, a three-tier defense framework for FL-based object detection, strengthening the spatial signature with the temporal signature analysis and uncertainty management. Extensive experiments with various adaptive adversaries show that \scheme{} can protect FL against all three model hijacking attacks and outperforms existing mitigation methods with high defense precision and a low false-positive rate. 

\noindent\textbf{Acknowledgments.} This research is partially sponsored by NSF grants 2302720, 2038029, 2026945, 1564097, an IBM faculty award, and a Cisco grant on Edge Computing. The first author acknowledges the IBM PhD Fellowship Award.

%%%%%%%%% REFERENCES
{\small
\bibliographystyle{ieee_fullname}
\bibliography{ref}
}

\end{document}